\documentclass[twocolumn,aps,pra,superscriptaddress,a4paper,sort&compress,balancelastpage]{revtex4-1}

\usepackage{amsmath}
\usepackage{amssymb}
\usepackage{gensymb}
\usepackage{bm}
\usepackage{graphicx}
\usepackage{float}

\usepackage[dvipsnames]{xcolor}
\usepackage{placeins}
\usepackage{braket}
\usepackage{bbold}
\usepackage{ulem}
\usepackage[colorlinks, linkcolor=blue, citecolor=blue, urlcolor=blue, breaklinks=red]{hyperref}

\newtheorem{theorem}{Theorem}

\newcommand{\SubFig}[2]{\ref{#1}{\color{blue}#2}}
\usepackage{bbm}
\newcommand{\1}{\mathbbm{1}}
\usepackage{lipsum}
\usepackage{physics}
\usepackage{dsfont}

\definecolor{Mycolor1}{HTML}{44aa99}
\definecolor{Mycolor2}{HTML}{cc6677}
\usepackage{orcidlink}

\begin{document}

\title{Quantum steering ellipsoids and quantum obesity in critical systems}

\author{Pedro Rosario~\!\!\orcidlink{0000-0002-7628-7373}}
\email{pedrorosario@estudante.ufscar.br}
\affiliation{Departamento de F\'isica, Universidade Federal de S\~{a}o Carlos, Rod.~Washington Lu\'is, km 235 - SP-310, 13565-905 S\~{a}o Carlos, SP, Brazil}

\author{Alan C. Santos~\!\!\orcidlink{0000-0002-6989-7958}}
\email{ac\_santos@df.ufscar.br}
\affiliation{Departamento de F\'isica, Universidade Federal de S\~{a}o Carlos, Rod.~Washington Lu\'is, km 235 - SP-310, 13565-905 S\~{a}o Carlos, SP, Brazil}

\begin{abstract}
Quantum obesity (QO) is new function used to quantify quantum correlations beyond entanglement, which also works as a witness for entanglement. Thanks to its analyticity for arbitrary state of bipartite systems, it represents an advantage with respect to other quantum correlations, like quantum discord for example. In this work we show that QO is a fundamental quantity to observe signature of quantum phase transitions. We also describe a mechanism based on local filtering operations able to intensify the critical behavior of the QO near to the transition point. To this end, we introduce a theorem stating how QO changes under local quantum operations and classical communications. This work opens perspective for the characterization of new phenomena in quantum critical systems through the analytically computable pairwise QO.
\end{abstract}


\maketitle


\section{Introduction}

Identifying quantum phase transitions (QPTs) is a key task for condensed-matter physics~\cite{sachdev_2011,SHOPOVA20031,Mott_1937,Mott_1949,Fisher:90,Sachdev:00,Fisher:99,Eddy:04,Anatoli:05,Kist:21}. Therefore, finding different ways to observe signatures of QPTs is a crucial role to characterize such phenomena. In this matter, quantum correlations have played a fundamental role in the characterization of QPT. Concurrence was initially used as a witness to investigate first order QPT for a certain class of one-dimensional magnetic systems, in which it was shown that entanglement can be employed within a framework of scaling theory~\cite{Osterloh:02}. Therefore, entanglement has been used to observe signature of transitions like Mott metal-insulator~\cite{Canella:21}, Superfluid-Insulator transitions in disordered systems~\cite{Canella:19,Canella:20}, phase ferro-antiferromagnetic transitions in magnetic systems~\cite{Osborne:02,Vidal:03,Verstraete:04,Wu:04}, among others~\cite{Gu:04,Larsson:05}. In addition, correlations beyond entanglement have been considered as indicators for QPT, including quantum discord~\cite{Sarandy:09,Dillenschneider:08}, Von-Neumann entropy~\cite{HUR20082208,Jia:08,Ivan:17,Jaime:22,KOPP20071466}, and mutual information~\cite{Wicks:07,Chen:10,Dong:21}. As a new quantity in this class of QPT witnesses, in this paper we show how \textit{quantum obesity} (QO)~\cite{Milne_2014} also can be used to detect QPTs.

\begin{figure}[t!]
	\includegraphics[width=\columnwidth]{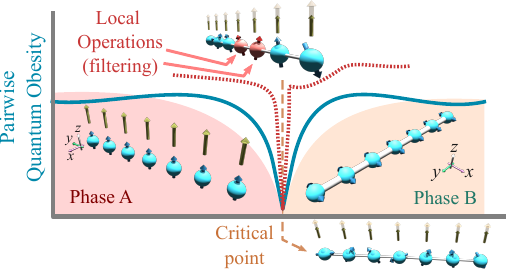}
	\caption{Pairwise QO used to detect the emergence of a QPT in a linear chain of spins. By driving the system from a phase A to a phase B, through a phase transition at zero temperature, the behavior of the QB will change at the transition point. Such a signature of QPT can be even more flashy if the system is submitted to local filtering operations before measuring pairwise QO for a pair of spins.}
	\label{Fig:Scheme}
\end{figure}

QO is a new class of quantum correlations beyond entanglement defined from quantum steering processes and the existence of steering ellipsoids~\cite{Jevtic_2014}, a geometrical object defined inside the local Bloch sphere of one of the parts of a bipartite system. These geometrical objects have been recently observed in an all-optical photonic qubit experimental setup~\cite{xu2023experimental}. For a given generic bipartite quantum state $\rho$, QO is defined by taking into account the volume of steering ellipsoids for the system. It can be interpreted as a kind of quantum correlation \textit{less restrict} than entanglement (a state may have QO without entanglement), but \textit{more restrict} than quantum discord~\cite{Milne_2014}. QO has been proposed as a figure of merit to study the entanglement swapping protocol analytically~\cite{Rosario_2023}. In this work we exploit QO as a detector for QPT. To this end, we study the behavior of the QO around a critical point for first order QPT in the Ising and XXZ Heisenberg spin chain in the thermodynamic limit. As a relevant previous work by Du \textit{et al.}~\cite{Du:21}, they showed how QPT is revealed through changes in the volume of the quantum steering ellipsoid. However, our results suggest that QO is a primitive for detecting QPT, and the results found in Ref.~\cite{Du:21} are consequences of the relation between QO and QPT, as sketched in Fig.~\ref{Fig:Scheme}. The Fig.~\ref{Fig:Scheme} also describes the last result of this work, in which the implementation of local filtering operations~\cite{Gisin:96,Verstraete:01,Verstraete:02} allow us to emphasize even more the unconventional behavior of the obesity around the critical point. 

\section{Preliminaries on two-qubit quantum steering and obesity}

In this section we review definitions and properties of quantum steering and obesity for bi-partite systems. As we will focus on phase transition for spin systems, our discussions concentrate on the two-qubit case, but further generalizations for quantum obesity in two-qudits states can be found in Ref.~\cite{Milne_2014}.

\subsection{Quantum obesity for two-qubit states}
In this section we introduce the general formulation of quantum obesity and some useful analytical solutions for different families of quantum states.
A general quantum state using the Bloch representation can be written as~\cite{Nielsen:Book}
\begin{align}
\hat{\varrho}^{\text{AB}}=\frac{1}{4}\sum_{i=0}^{3}\sum_{j=0}^{3}\mathcal{R}_{ij}\hat{\sigma_{i}}^{\text{A}}\otimes \hat{\sigma}^{\text{B}}_{j} , \label{eq:General_state}
\end{align}
where $\hat{\sigma}_{0}=\1$ and  $\{\hat{\sigma}_{i}\}^{3}_{i=1}$ define the Pauli operators. The correlations (quantum and classical) of a two-qubit state are revealed from the knowledge of a matrix $\mathcal{R}$, with matrix elements $\mathcal{R}_{ij}=\langle \hat{\sigma}_{i}\otimes \hat{\sigma}_{j}\rangle$, given by \cite{Gamel_2016}
\begin{align}
\mathcal{R}=\begin{pmatrix}
1 & a_{1} & a_{2}& a_{3}\\
b_{1} &\mathcal{T}_{11}&\mathcal{T}_{12}&\mathcal{T}_{13} \\
b_{2}&\mathcal{T}_{21}  & \mathcal{T}_{22}& \mathcal{T}_{23}\\
b_{3}& \mathcal{T}_{31}  & \mathcal{T}_{32} & \mathcal{T}_{33}
\end{pmatrix}=\begin{pmatrix}
1 & \vec{a}\\
\vec{b}^{T} &\mathcal{T}
\end{pmatrix} .
\end{align}
with $\vec{a} = [a_{1},a_{2},a_{3}]$ and $\vec{b} = [b_{1},b_{2},b_{3}]$ the local Bloch vectors for the part A and B, respectively, given by 
\begin{subequations}\label{eq:BlochVectors}
\begin{align}
\vec{a} &= [ \mathrm{tr}(\hat{\varrho} \hat{\sigma}_{1}^{A})\otimes\hat{\1}^{B}, \mathrm{tr}(\hat{\varrho} \hat{\sigma}_{2}^{A}\otimes\hat{\1}^{B}), \mathrm{tr}(\hat{\varrho} \hat{\sigma}_{3}^{A}\otimes\hat{\1}^{B})] ,  \\
\vec{b} &= [ \mathrm{tr}(\hat{\varrho} \hat{\1}^{A}\otimes\hat{\sigma}_{1}^{B}), \mathrm{tr}(\hat{\varrho} \hat{\1}^{A}\otimes\hat{\sigma}_{2}^{B}), \mathrm{tr}(\hat{\varrho} \hat{\1}^{A}\otimes\hat{\sigma}_{3}^{B})] .
\end{align} 
\end{subequations}

The block matrix $\mathcal{T}$ is particularly relevant because its set of singular values provide information to compute quantum correlations as CHSH-nonlocality~\cite{Horodecki_1995_CHSH} and EPR-steering~\cite{Costa_2016}, as well as to predict performance of quantum tasks like the teleportation of unknown states~\cite{Horodecki:96}. Beyond other quantum correlations, the correlation matrix $\mathcal{R}$ as a whole is used to define the so called {\it quantum obesity} for two-qubits, belonging to the Hilbert space $\mathcal{H}_{2} = \mathbf{C}^{2}\otimes \mathbf{C}^{2}$. QO is formally defined through the determinant of the matrix $\mathcal{R}$ as~\cite{Milne_2014}
\begin{align}
\Omega=\vert\mathrm{det}(\mathcal{R})\vert^{1/4} .
\label{eq:Obesity}
\end{align}

Although the QO can be analytically computed, to be pragmatic we will focus on three families of states of interest to the phenomenon we shall investigate in this work. The class of states we are interested can be written in a unified way as
\begin{align}
\hat{\varrho}_{(k)} =\begin{pmatrix}
\varrho_{11} & \delta_{k3}\varrho_{12} & \delta_{k2}\varrho_{13} & \varrho_{14}\\
\delta_{k3}\varrho_{21} &\varrho_{22}&\varrho_{23}& \delta_{k2}\varrho_{24} \\
\delta_{k2}\varrho_{31}&\varrho_{32}  & \varrho_{33} & \delta_{k3}\varrho_{34}\\
\varrho_{41}& \delta_{k2}\varrho_{42}  & \delta_{k3}\varrho_{43} & \varrho_{44}
\end{pmatrix} , \label{eq:DensityMatrixGeneric}
\end{align}
where $\delta_{kn}$ is the Kronecker's delta and $k =\{1,2,3\}$. Notice that, except for the state $\hat{\varrho}_{(1)}$, $\hat{\varrho}_{(k)}$ cannot be written as a $X$-state (which includes Bell Diagonal states). For this class of states, one finds that the QO takes the simply form
\begin{align}
\Omega=2\abs{ \big(|\varrho_{23}|^{2}-|\varrho_{14}|^{2}\big)\big(\varrho_{22}\varrho_{33}-\varrho_{11}\varrho_{44}\big)}^{1/4} .
\label{eq:Obesity_X}
\end{align}

This result will be used throughout our manuscript to study the signature of phase transitions via QO.

The geometrical picture of QO, justifying then such a terminology, can be seen from the \textit{steering ellipsoid} associated to the quantum state under consideration. Let $\hat{\varrho}^{\text{AB}}$ be a general two qubit state as defined in Eq.~\eqref{eq:General_state}, then local measurements and classical communication (LOCC) in the part B will `steer' the Bloch sphere of the subsystem A to a new geometry equivalent to an ellipsoid~\cite{Jevtic_2014}. In other words, consider $\hat{\varrho}^{\text{A}}_{\text{steer}}=\text{Tr}_{B}(\hat{\varrho}^{\text{AB}}_{\text{LOCC}})$ as the reduced density matrix of A after local measurements in B, with $\hat{\varrho}^{\text{AB}}_{\text{LOCC}} = \text{LOCC}[\hat{\varrho}^{\text{AB}}]$ the density matrix after the LOCC in B. So, any arbitrary state $\hat{\varrho}^{\text{A}}_{\text{steer}}$ lives inside an ellipsoidal region in the Block sphere of A described by a vector $\vec{c}_{A}$ and a matrix $\mathcal{Q}_{A}$ given, respectively, by
\begin{align}
\vec{c}_{A}&= \gamma_{b} \left(\vec{a}-\mathcal{T}\cdot\vec{b}\right) ,\\
\mathcal{Q}_{A}&=\gamma_{b}\left(\mathcal{T}-\vec{a}\cdot\vec{b}^{T}\right)\left(\1 +\gamma_{b}\vec{b}\cdot\vec{b}^{T}\right)\left(\mathcal{T}^{T}-\vec{b}\cdot\vec{a}^{T}\right),
\end{align}
where $\vec{c}_{A}$ localizes the center of an ellipsoid inside the local Bloch sphere of the sub-space $\text{A}$, and the eigenvalues $\{q_{i}\}_{i=1}^{3}$ and eigenvector of $\mathcal{Q}_{A}$ define the semiaxes lengths $s_{i}=\sqrt{q_{i}}$ and ellipsoid orientation, respectively. One defines the parameter $\gamma_{b} = 1/(1-|\vec{b}|^{2})$, assuming $|\vec{b}|<1$. Remarkably, by taking in to account the QO of $\hat{\varrho}^{\text{AB}}$, it is possible to show that the volume $\mathcal{V}_{A}$ of such a steering ellipsoid is given by~\cite{Milne_2014} 
\begin{align}
\mathcal{V}_{A}=\frac{4\pi}{3}\gamma^{4}_{b}\Omega^{4} . \label{eq:Vol}
\end{align}

It is worth to mention that, although the quantity $\mathcal{V}_{A}$ is a purely mathematical definition, its applicability as a witness of entanglement~\cite{Jevtic_2014} and to quantum phase transitions~\cite{Du:21} suggests such a definition may have a deep interpretation in physical realm. On the other hand, it was shown that QO $\Omega$ represents an upper bound for concurrence~\cite{Milne_2014}.

\section{Quantum Obesity and quantum phase transitions}

In order to make a connection of the Eq.~\eqref{eq:Vol} as a witness of QPT, it is reasonable to assume that the reduced density matrix for the system depends on a given parameter $\xi$ used to characterize the phase transition. Mathematically, we write $\gamma_{b} \equiv \gamma_{b}(\xi)$ and $\Omega\equiv \Omega(\xi)$, such that, $\mathcal{V}_{A}\equiv \mathcal{V}_{A}(\xi)$. Therefore, we have for a given parameter $\xi$
\begin{align}
\partial_{\xi}\mathcal{V}_{A}(\xi)  = \frac{16\pi \gamma^{3}_{b}(\xi)\Omega^{3}(\xi)}{3} \big[ \Omega(\xi) \partial_{\xi}\gamma_{b}(\xi) + \gamma_{b}(\xi) \partial_{\xi}\Omega(\xi) \big] . \label{eq:DerivateVolume}
\end{align}

The above result allows us to conclude that abrupt changes (discontinuity) in the function $\partial_{\xi}\mathcal{V}_{A}(\xi)$ only comes from one of the terms $\partial_{\xi}\gamma_{b}(\xi)$ or $\partial_{\xi}\Omega(\xi)$ (or from both). It leads to our main result: \textit{the derivative of QO, $\partial_{\xi}\Omega(\xi)$, is a new quantity able to reveal signatures of first-order QPT}. In this section we exemplify such a conjecture with application to two different phase transitions. First, we study the QPT for the Ising chain under a transverse field in the thermodynamic limit. As a last example, we briefly discuss about the results discussed in Ref.~\cite{Du:21}, the ideal scenario to introduce and exemplify the second main result of this work.

\subsection{The Ising-Lenz chain in thermodynamic limit}

The Hamiltonian that describes the one-dimensional chain of $N$ spin-$\frac{1}{2}$ in a transverse $Z$-field may be given by the Ising-Lenz model as
\begin{align}
\hat{H}_{\mathrm{IL}}(\lambda) = - \sum_{i=1}^{N}(\lambda \hat{\sigma}^{x}_{i}\hat{\sigma}^{x}_{i+1}+ \omega_{0}\hat{\sigma}^{z}_{i}),
\end{align}
introduced by de Gennes to describe ferroelectric crystals~\cite{deGennes:63}. We assume the transverse field along the quantization $Z$-axes defining the spin up ($\ket{\uparrow}$) and spin down states ($\ket{\downarrow}$), such that $\hat{\sigma}^{z}\ket{\uparrow}=\ket{\uparrow}$ and $\hat{\sigma}^{z}\ket{\downarrow}=-\ket{\downarrow}$. Since we are interested in the quantum nature of a phase transition, we assume this system is brought to the thermal equilibrium with a reservoir at zero temperature ($T=0$). In this way, the thermal state of the system reads ($k_{B}$ the Boltzmann constant)
\begin{align}
\hat{\rho}_{\mathrm{th}}(\lambda) = \lim_{T \rightarrow 0} \left[\frac{e^{- \hat{H}_{\mathrm{IL}}(\lambda)/k_{B} T}}{\mathcal{Z}(\lambda,T)}\right] = \hat{\rho}_{\mathrm{gr}}(\lambda) ,
\end{align}
with $\mathcal{Z}(\lambda,T) = \mathrm{tr}\big(\exp\small(- \hat{H}_{\mathrm{IL}}(\lambda)/k_{B} T\small) \big)$ the partition function of the system, and $\hat{\rho}_{\mathrm{gr}}(\lambda)=\ket{E_{\mathrm{gr}}(\lambda)}\bra{E_{\mathrm{gr}}(\lambda)}$ the ground state of $\hat{H}_{\mathrm{IL}}(\lambda)$. In this way we can focus on the study of the pairwise QO of the ground state of $\hat{H}_{\mathrm{IL}}(\lambda)$ as a function of the parameter $\lambda$. This model presents a critical point at $\lambda = \lambda_{\mathrm{c}} = \omega_{0}$. So, we investigate the phase transition by varying the parameter in the interval $\lambda \in [0,2\lambda_{\mathrm{c}}]$. When $\lambda = 0$ the spins get aligned along $Z$-direction in the classical paramagnet state $\ket{\uparrow\uparrow\cdots\uparrow\uparrow}$. When $\lambda \gg \lambda_{\mathrm{c}}$ the collective spin states point in $X$-direction, where the system state converges to a superposition of its two degenerate ground states $\ket{+\cdots +}$ and $\ket{-\cdots -}$~\cite{Sarandy:09}, where $\ket{\pm} = (1/\sqrt{2})(\ket{\uparrow}\pm \ket{\downarrow})$. For an arbitrary value of $\lambda$ in the thermodynamic limit, the thermal ground state is analytically found~\cite{Barouch:70,Barouch:71}. Since we are interested in pairwise QO, we focus on reduced density matrix for an arbitrary pair of spins $(i,j)$ as~\cite{Barouch:70,Barouch:71,Osborne:02,Dillenschneider:08,Maziero:10}
\begin{align}
\hat{\varrho}_{i,j} (\lambda) = \begin{pmatrix}
\mathcal{A}^{+} & 0 & 0& \mathcal{C}_{i,j}^{-}\\
0 &\mathcal{B}_{i,j}&\mathcal{C}_{i,j}^{+}&0 \\
0&\mathcal{C}_{i,j}^{+} & \mathcal{B}_{i,j} & 0\\
\mathcal{C}_{i,j}^{-}& 0  & 0 & \mathcal{A}^{-}
\end{pmatrix}
\label{eq:Normal_state}
\end{align}
where $\mathcal{A}^{\pm}=(1/4)\pm\langle \hat{\sigma}^{z} \rangle/2$, $\mathcal{B}_{i,j}=(1-\langle \hat{\sigma}^{z}_{i}\hat{\sigma}^{z}_{j}\rangle)/4$ and $\mathcal{C}_{i,j}^{\pm}=(\langle \hat{\sigma}^{x}_{i}\hat{\sigma}^{x}_{j}\rangle \pm \langle \hat{\sigma}^{y}_{i}\hat{\sigma}^{y}_{j}\rangle)/4$, with the expected values
\begin{subequations}
\begin{align}
&\langle \hat{\sigma}^{z}\rangle=-\frac{1}{\pi}\int_{0}^{\pi}d\phi \frac{(1+\cos{\phi})}{\omega_{\phi}},\\
&\langle \hat{\sigma}^{x}_{i}\hat{\sigma}^{x}_{i+k}\rangle=\begin{vmatrix}
G_{-1} & G_{-2} & \hdots& G_{-k}\\
G_{0} &G_{-1}&\hdots&G_{-k+1} \\
\vdots&\vdots  & \ddots & \vdots\\
G_{k-2}& G_{k-3}  & \hdots & G_{-1}
\end{vmatrix} ,\\
&\langle \hat{\sigma}^{y}_{i}\hat{\sigma}^{y}_{i+k}\rangle=\begin{vmatrix}
G_{1} & G_{0} & \hdots& G_{-k+2}\\
G_{2} &G_{1}&\hdots&G_{-k+3} \\
\vdots&\vdots  & \ddots & \vdots\\
G_{k}& G_{k-1}  & \hdots & G_{1}
\end{vmatrix} , \\
&\langle \hat{\sigma}^{z}_{i}\hat{\sigma}^{z}_{i+k}\rangle =\langle \hat{\sigma}^{z}\rangle^{2}-G_{k}G_{-k},
\end{align}
\end{subequations}
where $\omega_{\phi}=\sqrt{(\lambda \sin{\phi})^{2}+(1+\lambda \cos{\phi})^{2}}$ and
\begin{align}
G_{\ell} =\int_{0}^{\pi}   \frac{d\phi}{\pi} \frac{1}{\omega_{\phi}}\Big[\cos(\ell\phi)+\lambda \cos((\ell+1)\phi)\Big] .
\end{align}

Therefore, by using the Eq.~\eqref{eq:Obesity_X}, one finds the QO for the state $\hat{\varrho}_{i,j}$ as
\begin{align}
\Omega_{i,j} (\lambda) =2\big|[(\mathcal{C}_{i,j}^{+})^{2}-(\mathcal{C}_{i,j}^{-})^{2}][(\mathcal{B}_{i,j})^{2}-\mathcal{A}^{+}\mathcal{A}^{-}]\big|^{1/4} .
\end{align}

\begin{figure}[t!]
	\includegraphics[width=\columnwidth]{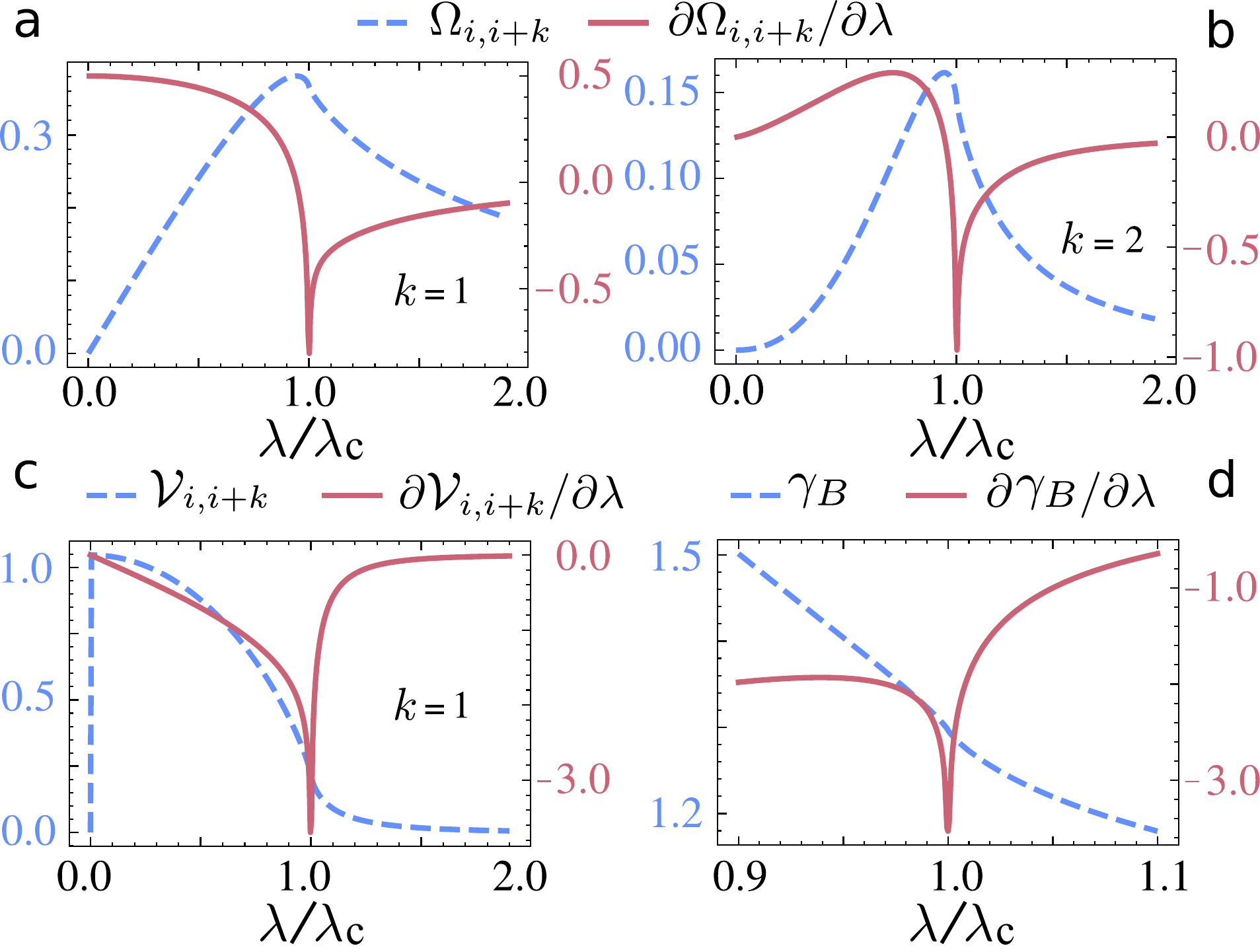}
	\caption{(a) QO for next neighbor $\Omega_{i,i+1}$, and (b) next-nearest neighbor $\Omega_{i,i+2}$ as function of the spin-spin coupling strength $\lambda$. At the critical point we observe the abrupt change in the derivative of the obesity, indicating a first order QPT. (c) Steering ellipsoid volume and its derivative for the nearest  neighbor spins. (d) Behavior of the function $\gamma_{b}$ and its derivative, the quantity $\partial_{\lambda} \gamma_{b} (\lambda)$ contributes to the volume discontinuity.}
	\label{Fig:Normal}
\end{figure}

We first study the behavior of the QO for two nearest neighbor spins $(i, i+1)$, and for the next-nearest neighbor spins $(i,i+2)$. Also, we investigate the derivative of the obesity $\partial_{\lambda}\Omega_{i,j} (\lambda)$ in order to observe signature of first-order QPT. The result is shown in Fig.~\ref{Fig:Normal}. The sharp change in $\partial_{\lambda}\Omega_{i,i+1} (\lambda)$ and $\partial_{\lambda}\Omega_{i,i+2} (\lambda)$, Figs.~\SubFig{Fig:Normal}{a} and~\SubFig{Fig:Normal}{b} respectively, are a reasonable signature of QPT at the known critical point $\lambda = \lambda_{\mathrm{c}}$. Therefore, we conclude that the obesity itself is a sufficient correlation for predicting a QPT behavior for this model.

However, given that the volume of the quantum steering ellipsoid depends on the obesity, as stated in Eq.~\eqref{eq:DerivateVolume}, a non-analytical behavior for the derivative in obesity will lead to a (quasi-)non-analytical behavior for the derivative of the volume at the exact critical value. In fact, from Eq.~\eqref{eq:Vol} the volume of the steering ellipsoid for the pair of spins $(i, j)$ is given by
\begin{align}
\mathcal{V}_{i,j}(\lambda)=\frac{4\pi}{3}\frac{\Omega^{4}_{i,j} (\lambda)}{(1-|\langle \hat{\sigma}^{z}\rangle|^{2})^{2}} ,
\end{align}
where we used $|\vec{b}|=|\mathcal{A}^{+}-\mathcal{A}^{-}| = |\langle \hat{\sigma}^{z}\rangle|$ for the state in Eq.\eqref{eq:Normal_state}. From this equation we have access to the graph shown in Fig.~\SubFig{Fig:Normal}{c} of the volume $\mathcal{V}_{i,i+1}(\lambda)$ for nearest neighbor spins, including also the curve for $\partial_{\lambda} \mathcal{V}_{i,i+1}(\lambda)$. It is worth highlighting that the fast decreasing in the function $\partial_{\lambda} \mathcal{V}_{i,i+1}(\lambda)$ around $\lambda = \lambda_{\mathrm{c}}$ contains a contribution of the coefficient $\partial_{\lambda} \gamma_{b} (\lambda)$. As $\gamma_{b} (\lambda)$ only depends on the magnetization of the system, the obesity is the quantity related to quantum correlations in the model. 

\subsection{Phase transitions in the $XXZ$ model}

In this section we introduce the second main result of our manuscript, where we exemplify our result through the phase transition in the XXZ (Heisenberg-Ising) model~\cite{Yang:660}. First of all, let us consider the obesity for the class of states known as Bell-Diagonal (BD) states. An arbitrary two-qubit state that can be described using only three parameters $\{c_{j}\}_{j=1}^{3}$ as follows 
\begin{align}
\hat{\varrho}^{\text{BD}}=\frac{1}{4}\left(\1+\sum \nolimits_{j=1}^{3}c_{j}\hat{\sigma}^{\text{A}}_{j}\otimes\hat{\sigma}^{\text{B}}_{j}\right) .
\label{eq:BDstate}
\end{align}

This family of states is a special case of the density matrix $\hat{\varrho}_{(1)}$ in Eq.~\eqref{eq:DensityMatrixGeneric} when $\varrho_{11}=\varrho_{44}$ and $\varrho_{22}=\varrho_{33}$. Consequently, following Eq.~\eqref{eq:Obesity_X} for Bell-diagonal the QO reads
\begin{align}
\Omega_{\text{BD}}=\abs{ c_{1}c_{2}c_{3}}^{1/4} .
\label{eq:Obesity_bellDiagonal}
\end{align}

Also, given the volume of the steering ellipsoid defined in Eq.~\eqref{eq:Vol}, we compute the parameter $\gamma_{b}$ from the local Bloch vector $\vec{b}$ in Eq.~\eqref{eq:BlochVectors} and we find all components of $\vec{b}$ are identically zero, and therefore $|\vec{b}|=0$, which leads to $\gamma_{b} = 1$ for the state $\hat{\varrho}^{\text{BD}}$. Therefore, here we highlight that the volume of the ellipsoid is
\begin{align}
\mathcal{V}_{\text{BD}} = \frac{4\pi}{3}\Omega_{\text{BD}}^{4} = \frac{4\pi}{3}\abs{ c_{1}c_{2}c_{3}}.
\label{eq:V_bellDiagonal}
\end{align}

The above equation is enough to introduce the second main result of this work. For systems with ground states mapped into two-spin Bell-diagonal states we have $\partial_{\xi}\mathcal{V}_{\text{BD}} \propto \partial_{\xi}\Omega_{\text{BD}}$, for any physical parameter $\xi$. As an immediate example of this result, let us consider the Hamiltonian describing the one-dimensional spin-$\frac{1}{2}$ chain with anisotropic nearest-neighbor interactions given by
\begin{align}
\hat{H}_{\mathrm{HI}}(\Delta) = \hbar J \sum_{i=1}^{N} \left(\hat{\sigma}^{x}_{i}\hat{\sigma}^{x}_{i+1} + \hat{\sigma}^{y}_{i}\hat{\sigma}^{y}_{i+1}+ \Delta \hat{\sigma}^{z}_{i}\hat{\sigma}^{z}_{i+1}\right) .
\end{align}

By using the solution for the ground state of the system~\cite{Yang:660}, based on the Bethe anzats~\cite{Yang:66,Yang:66II}, Du \textit{et al}~\cite{Du:21} reported a geometrical interpretation of a signature of QPT for this model. It can be done by computing the reduced density matrix $\hat{\rho}_{i,j}^{\mathrm{HI}}(\Delta)$, for a pair of spins $(i,j)$ of the thermal ground of the model. For this state, Du \textit{et al} showed that its ellipsoid volume has a discontinuity at the transition point $\Delta = -1$. At this point the system goes from the ferromagnetic ($\Delta < -1$) to the gapless phase ($-1 < \Delta < 1$) through a transition point. Because reduced density matrix $\hat{\rho}_{i,j}^{\mathrm{HI}}(\Delta)$ can be written from Eq.~\eqref{eq:BDstate} under the mapping
\begin{align}
c_{1} = \langle \hat{\sigma}^{z}_{i}\hat{\sigma}^{z}_{i+1} \rangle , ~~ c_{2} = c_{3} = \langle \hat{\sigma}^{x}_{i}\hat{\sigma}^{x}_{i+1} \rangle + \langle \hat{\sigma}^{y}_{i}\hat{\sigma}^{y}_{i+1} \rangle .
\end{align}

In this model, from Eq.~\eqref{eq:V_bellDiagonal}, we then conclude that $\partial_{\Delta}\gamma_{b} = 0$, and therefore the signature of phase transition observed through quantum steering ellipsoid in Ref.~\cite{Du:21} comes \textit{fundamentally from the QO}, which quantifies quantum correlations beyond entanglement. Furthermore, this result also leads to the conclusion that any signature of high-orders phase transitions~\cite{Wu:04} also comes genuinely from the QO. 

\section{Phase transition signature after local filtering operations}

In this section we present the last result of this work, by showing how to use \textit{local filtering operations} (LFOs)~\cite{Gisin:96,Verstraete:01,Verstraete:02} to intensify the signature of phase transition through QO. Briefly, LFO is a procedure through which the LOCC is used to ``project" an arbitrary two-qubit state, in Eq.~\eqref{eq:General_state}, on the space of states constituted by Bell-diagonal states. After this projection, or filtering, the hidden amount of correlations in the system can be revealed by applying local operators of the form $\mathcal{O}_{A}\otimes \mathcal{O}_{B}$~\cite{Gisin:96}. Therefore, given an arbitrary two-spin state $\hat{\rho}_{i,j}$ the filtered state reads
\begin{align}
\hat{\varrho}^{F}_{i,j}=\frac{(\mathcal{O}_{i}\otimes \mathcal{O}_{j})\hat{\rho}_{i,j}(\mathcal{O}_{i}\otimes \mathcal{O}_{j})}{\text{tr}[(\mathcal{O}_{i}\otimes \mathcal{O}_{j})\hat{\rho}_{i,j}(\mathcal{O}_{i}\otimes \mathcal{O}_{j})]} .
\label{eq:Filter_state}
\end{align}

If the operation $\mathcal{O}_{i}\otimes \mathcal{O}_{j}$ is properly optimized, then $\hat{\varrho}^{F}_{i,j}$ gets arbitrarily close to a Bell-diagonal state~\cite{Verstraete:01,Verstraete:02}. 

\begin{theorem}[Filtered Quantum Obesity] \label{Theorem1}
Consider a quantum state $\hat\rho$ of a system $AB$ composed by two qubits, and its corresponding state $\hat\rho^{F}$ after local filtering operations given by $\hat{\mathcal{O}} = \hat{\mathcal{O}}_{A}\otimes\hat{\mathcal{O}}_{B}$. By denoting the QO of $\hat\rho$ as $\Omega[\hat\rho]$, then the obesity of the filtered density matrix $\hat\rho^{F}$ will be
\begin{align}
\Omega[\hat\rho^{F}] = \Omega[\hat\rho] \mathcal{F}_{\hat\rho,\hat{\mathcal{O}}} ,
\end{align}
with $\mathcal{F}_{\hat\rho,\hat{\mathcal{O}}} = |\mathrm{tr}(\hat{\mathcal{O}}\hat\rho\hat{\mathcal{O}}^{\dagger})|^{-1}$ the filtering function of the state $\hat\rho$ given the operation $\hat{\mathcal{O}}$.
\end{theorem}

\begin{figure}[t!]
	\centering
	\includegraphics[width=\columnwidth]{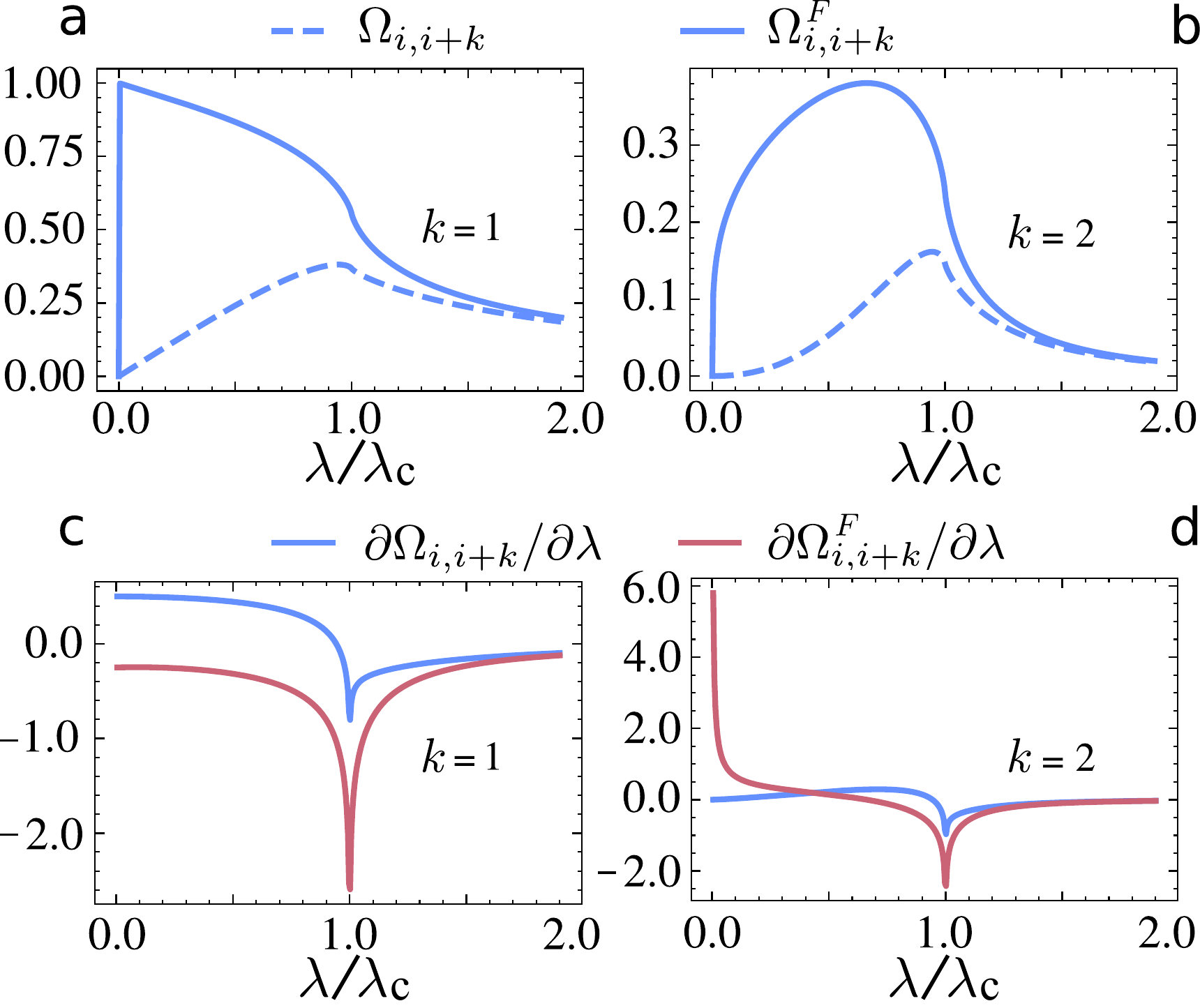}
	\caption{The QO for the system and its corresponding behavior after applying FLOs, for the case of (a) nearest neighbor and (b) next-nearest neighbor spins. (c) and (d) are the derivative of the obesity and its filtered counterpart for (c) nearest neighbor and (d) next-nearest neighbor spins.}
	\label{Fig:Filter}
\end{figure}

One of the main consequences of this theorem applied to QPTs is the possibility to amplify the discontinuity effect of the obesity at the critical point of a first order phase transition. In fact, consider the obesity for the two-qubit reduced density matrix $\rho_{i,j}(\xi)$ and its obesity $\Omega_{i,j}(\xi)$, with $\xi$ the free parameter. Also, by denoting $\Omega_{i,j}^{F}(\xi)$ the obesity of its filtered counterpart $\rho_{i,j}^{F}(\xi)$, we have
\begin{align}
	\partial_{\xi} \Omega_{i,j}^{F}(\xi) = \mathcal{F}_{\hat\rho_{i,j},\hat{\mathcal{O}}}(\xi) \partial_{\xi} \Omega_{i,j}(\xi) + \Omega_{i,j}(\xi) \partial_{\xi} \mathcal{F}_{\hat\rho_{i,j},\hat{\mathcal{O}}}(\xi) .
\end{align}

Therefore, after the filtering operations, any discontinuity of the function $\partial_{\xi} \Omega_{i,j}(\xi)$ can be not only preserved, but it also can be intensified due to the derivative in the filtering function $\partial_{\xi} \mathcal{F}_{\hat\rho_{i,j},\hat{\mathcal{O}}}(\xi)$. To exemplify such a result, let us now consider the Ising-Lenz chain as an example of the above discussion. For such a model, the optimal operators for the obesity filtering are given by
\begin{align}
\mathcal{O}_{i} = \mathcal{O}_{j} = \begin{bmatrix}
\eta & 0 \\
0 & 1
\end{bmatrix} ,
\end{align}
where $\eta=(\mathcal{A}^{-}/\mathcal{A}^{+})^{1/4}$, and it satisfies $0\leq \eta^{2}\leq 1 $. After such a filtering given by the above equation it is possible to show that the QO of the output filtered state $\hat{\varrho}^{F}_{i,j}$ reads
\begin{align}
\Omega^{F}_{i,j}(\lambda)=\mathcal{F}_{i,j}(\lambda)\Omega_{i,j}(\lambda)
\end{align}
where $\mathcal{F}_{i,j}(\lambda)=|(\mathcal{B}_{i,j}+\sqrt{\mathcal{A}^{+}\mathcal{A}^{-}})|^{-1/2}/\sqrt{2}$. Consequently, we find the filtered obesity as
\begin{align}
\partial_{\lambda} \Omega^{F}_{i,j}(\lambda) = \Omega_{i,j}(\lambda)\partial_{\lambda} \mathcal{F}_{i,j}(\lambda) + \mathcal{F}_{i,j}(\lambda)\partial_{\lambda} \Omega_{i,j}(\lambda) \label{eq:FilterDerivate},
\end{align}
where $\mathcal{F}_{i,j}(\lambda) \neq 0$ in the interval $\lambda \in [0,2\lambda_{\mathrm{c}}]$.

In Figs.~\SubFig{Fig:Filter}{a} and~\SubFig{Fig:Filter}{b} we observe the immediate application of the Theorem~\ref{Theorem1}, where the filtering operations reveals quantum correlations than the non-filtered counterpart. It is important to highlight here the amount of hidden correlation revealed by the filtering for $\lambda$ slightly different from zero, as shown by the abrupt change of the obesity for the filtered state. To better understand this sudden change in the obesity, it is worth stressing that before the local filtering the system is at the thermal equilibrium with a bath at zero temperature. So, such a sudden change in the obesity is mainly created by the local measurement which brings the system out-of-equilibrium with its environment. In addition, the derivative for the system obesity and its filtered counterpart are depicted in Figs.~\SubFig{Fig:Filter}{c} and~\SubFig{Fig:Filter}{d}. As already predicted in Eq.~\eqref{eq:FilterDerivate}, it is possible to see an ``enhancement" in the amplitude of the derivative of the obesity at the transition when we apply the local filter.

Although mathematically intuitive, from Eq.~\eqref{eq:FilterDerivate}, it is particularly intriguing that the signature of the QPT is observed even after applying the local filtering procedure. We mean, the phase transition is observed for the equilibrium state, but its signature will be observed (and some time amplified) when applying local operations that bring the system far from the equilibrium state. Moreover, we also can argue that a similar result would be expected for concurrence used as witness for QPT. In fact, given the filtered concurrence~\cite{Verstraete:01} also obeys a theorem similar to the Theorem~\ref{Theorem1}, we also expect such an ``enhancement" can be obtained if we use pairwise entanglement to characterize a QPTs.


\section{Conclusion}

The conclusions of this manuscript are threefold. First, we conjecture and provide examples showing that correlations beyond entanglement created in the system as QO are genuine identifiers for QPTs. Also, as a consequence of this result we state that the geometrical picture~\cite{Du:21} of the QPT comes from QO response to the phase transition. Second, we introduced a new theorem that rules the change in the QO after applying LFOs. Such a theorem is then used to support the last finding of this work: the signature of QPT through QO becomes even more evident after applying local filtering at the ground thermal state of the system before measuring the correlations in the system. These results are used in two different models of one-dimensional spin-$\frac{1}{2}$ with nearest neighbor, namely, the Ising-Lenz and the XXZ Heisenberg chains in the thermodynamic limit.

This works pave the way to further studies and development of mechanisms based on QO to study of QPT in different systems. For example, the generalization of our approach to take into account pairwise QO in qudits system would be the immediate path forward for studies of QPT in systems like Bose-Hubbard model~\cite{Gersch:63}, spin-$1$ bosons in optical lattices~\cite{Tsuchiya:04}, or ultracold bosons and fermions in an optical lattices (Bose-Fermi Hubbard Hamiltonian)~\cite{Albus:03}. The proposals of this work have potential to applications beyond the content of condensed-matter physics and QPT. For example, we expect the Theorem~\ref{Theorem1}, as well as its generalized version to qudits, will be applied to new progressions in the field of quantum information and communication theory.

\begin{acknowledgements}
	This work is supported by the São Paulo Research Foundation (FAPESP) (Grants 2019/22685-1 and 2022/12382-4).
\end{acknowledgements}

\appendix

\section{Demonstration of the Theorem~\ref{Theorem1}}\label{Appendix:ProofTheorem}

Let a general two-qubit state
\begin{align}
	\hat{\varrho}^{\text{AB}}=\sum_{i=0}^{3}\sum_{j=0}^{3}\mathcal{R}_{ij}\hat{\sigma}_{i}^{\text{A}}\otimes \hat{\sigma}_{j}^{\text{B}} .
	\label{AP:eq:App1}
\end{align}

In this representation, we can define a matrix $\hat{\mathcal{R}}$ with elements given by $\mathcal{R}_{ij}$, such that information about the quantum state $\hat{\varrho}^{\text{AB}}$ can be obtained from $\mathcal{R}_{ij}$. By applying local operations on part $\text{A}$ and $\text{B}$, the filtered state $\hat{\varrho}^{\text{AB}}$ 
\begin{align}
	\left(\hat{\varrho}^{\text{AB}}\right)^{F}=\frac{(\hat{\mathcal{O}}_{A}\otimes \hat{\mathcal{O}}_{B})\hat{\varrho}^{\text{AB}}(\hat{\mathcal{O}}_{A}^{\dagger}\otimes \hat{\mathcal{O}}_{B}^{\dagger})}{\text{Tr}[(\hat{\mathcal{O}}_{A}\otimes \hat{\mathcal{O}}_{B})\hat{\varrho}^{\text{AB}}(\hat{\mathcal{O}}_{A}^{\dagger}\otimes \hat{\mathcal{O}}_{B}^{\dagger})]} ,
	\label{AP:eq:Filter_state}
\end{align}
where are the local filters $\hat{\mathcal{O}}_{A},\hat{\mathcal{O}}_{B}$. Analogously to the strategy done in Ref.~\cite{Verstraete:01} to study the concurrence after local filtering, the new filtered density matrix $\mathcal{R}^{F}$ reads
\begin{align}
	\hat{\mathcal{R}}^{F}=\frac{\hat{L}_{A}\hat{\mathcal{R}}L^{T}_{B}}{\text{Tr}\left[(\hat{\mathcal{O}}_{A}^{\dagger}\hat{\mathcal{O}}_{A}\otimes \hat{\mathcal{O}}_{B}^{\dagger}\hat{\mathcal{O}}_{B})\hat{\varrho}^{\text{AB}}\right]} ,
	\label{AP:eq:R_filter}
\end{align}
where
\begin{align}
	\hat{L}_{A}= \frac{\hat{\Lambda}(\hat{\mathcal{O}}_{A}\otimes \hat{\mathcal{O}}_{A}^{*})\hat{\Lambda}^{\dagger}}{|\det \hat{\mathcal{O}}_{A} |}, ~~
	\hat{L}_{B}=\frac{ \hat{\Lambda}(\hat{\mathcal{O}}_{B}\otimes \hat{\mathcal{O}}_{B}^{*})\hat{\Lambda}^{\dagger}}{|\det \hat{\mathcal{O}}_{B} |} ,
\end{align}
with
\begin{align}
	\hat{\Lambda}=\frac{1}{\sqrt{2}}\begin{pmatrix}
		1 & 0 & 0&1\\
		0 &1&1&0\\
		0 &i & -i &0\\
		1& 0 & 0& -1
	\end{pmatrix} .
\end{align}

Therefore, the obesity for the filtered state will be
\begin{align}
\Omega^{F} &= |\det \mathcal{R}^{F}|^{1/4} = \frac{\abs{\det(\hat{L}_{A}\hat{\mathcal{R}}L^{T}_{B})}^{1/4}}{\abs{\mathrm{tr}\left[(\hat{\mathcal{O}}_{A}^{\dagger}\hat{\mathcal{O}}_{A}\otimes \hat{\mathcal{O}}_{B}^{\dagger}\hat{\mathcal{O}}_{B})\hat{\varrho}^{\text{AB}}\right]}^{1/4}} \nonumber \\
&= \frac{|\det\hat{\mathcal{R}}|^{1/4}}{\abs{\mathrm{tr}\left[(\hat{\mathcal{O}}_{A}^{\dagger}\hat{\mathcal{O}}_{A}\otimes \hat{\mathcal{O}}_{B}^{\dagger}\hat{\mathcal{O}}_{B})\hat{\varrho}^{\text{AB}}\right]}^{\dim \mathcal{H}_{AB}/4}}
\end{align}
where we used that $\mathrm{det}(\hat{L}_{A}\hat{\mathcal{R}}L^{T}_{B})=\det\hat{L}_{A} \det\hat{\mathcal{R}} \det L^{T}_{B}$ and 
\begin{align}
    \nonumber \det \hat{L}_{A/B}&=\frac{\det{\hat{\Lambda}(\hat{\mathcal{O}}_{A/B}\otimes \hat{\mathcal{O}}_{A/B}^{*})\hat{\Lambda}^{\dagger}}}{|\det \hat{\mathcal{O}}_{A/B} |^{4}}\\
    &\nonumber=\frac{(\det \hat{\mathcal{O}}_{A/B})^{2}(\det \hat{\mathcal{O}}_{A/B}^{*})^{2}}{|\det \hat{\mathcal{O}}_{A/B} |^{4}}=1
\end{align}

We also used that for any square matrix $\hat{X}$, we have $\mathrm{det}(c \hat{X}) = c^{\dim \hat{X}} \det \hat{X}$, with $\dim \hat{X}$ the dimension of $\hat{X}$. So, by defining the filtering function $\mathcal{F} = \abs{\mathrm{tr}\big[(\hat{\mathcal{O}}_{A}^{\dagger}\hat{\mathcal{O}}_{A}\otimes \hat{\mathcal{O}}_{B}^{\dagger}\hat{\mathcal{O}}_{B})\hat{\varrho}^{\text{AB}}\big]}^{- 1}$, we conclude that
\begin{align}
	\Omega^{F} &= \Omega \mathcal{F} .
\end{align}


\begin{thebibliography}{53}%
	\makeatletter
	\providecommand \@ifxundefined [1]{%
		\@ifx{#1\undefined}
	}%
	\providecommand \@ifnum [1]{%
		\ifnum #1\expandafter \@firstoftwo
		\else \expandafter \@secondoftwo
		\fi
	}%
	\providecommand \@ifx [1]{%
		\ifx #1\expandafter \@firstoftwo
		\else \expandafter \@secondoftwo
		\fi
	}%
	\providecommand \natexlab [1]{#1}%
	\providecommand \enquote  [1]{``#1''}%
	\providecommand \bibnamefont  [1]{#1}%
	\providecommand \bibfnamefont [1]{#1}%
	\providecommand \citenamefont [1]{#1}%
	\providecommand \href@noop [0]{\@secondoftwo}%
	\providecommand \href [0]{\begingroup \@sanitize@url \@href}%
	\providecommand \@href[1]{\@@startlink{#1}\@@href}%
	\providecommand \@@href[1]{\endgroup#1\@@endlink}%
	\providecommand \@sanitize@url [0]{\catcode `\\12\catcode `\$12\catcode
		`\&12\catcode `\#12\catcode `\^12\catcode `\_12\catcode `\%12\relax}%
	\providecommand \@@startlink[1]{}%
	\providecommand \@@endlink[0]{}%
	\providecommand \url  [0]{\begingroup\@sanitize@url \@url }%
	\providecommand \@url [1]{\endgroup\@href {#1}{\urlprefix }}%
	\providecommand \urlprefix  [0]{URL }%
	\providecommand \Eprint [0]{\href }%
	\providecommand \doibase [0]{http://dx.doi.org/}%
	\providecommand \selectlanguage [0]{\@gobble}%
	\providecommand \bibinfo  [0]{\@secondoftwo}%
	\providecommand \bibfield  [0]{\@secondoftwo}%
	\providecommand \translation [1]{[#1]}%
	\providecommand \BibitemOpen [0]{}%
	\providecommand \bibitemStop [0]{}%
	\providecommand \bibitemNoStop [0]{.\EOS\space}%
	\providecommand \EOS [0]{\spacefactor3000\relax}%
	\providecommand \BibitemShut  [1]{\csname bibitem#1\endcsname}%
	\let\auto@bib@innerbib\@empty
	\bibitem [{\citenamefont {Sachdev}(2011)}]{sachdev_2011}%
	\BibitemOpen
	\bibfield  {author} {\bibinfo {author} {\bibfnamefont {S.}~\bibnamefont
			{Sachdev}},\ }\href {\doibase 10.1017/CBO9780511973765} {\textit {\bibinfo
			{title} {Quantum Phase Transitions}}},\ \bibinfo {edition} {2nd}\ ed.\
	(\bibinfo  {publisher} {Cambridge University Press},\ \bibinfo {year}
	{2011})\BibitemShut {NoStop}%
	\bibitem [{\citenamefont {Shopova}\ and\ \citenamefont
		{Uzunov}(2003)}]{SHOPOVA20031}%
	\BibitemOpen
	\bibfield  {author} {\bibinfo {author} {\bibfnamefont {D.~V.}\ \bibnamefont
			{Shopova}} and \bibinfo {author} {\bibfnamefont {D.~I.}\ \bibnamefont
			{Uzunov}},\ }\enquote{\bibinfo {title} {Some basic aspects of quantum phase
			transitions}},\ \href {\doibase
		https://doi.org/10.1016/S0370-1573(03)00076-0} {\bibfield  {journal}
		{\bibinfo  {journal} {Physics Reports}\ }\textbf {\bibinfo {volume} {379}},\
		\bibinfo {pages} {1} (\bibinfo {year} {2003})}\BibitemShut {NoStop}%
	\bibitem [{\citenamefont {Mott}\ and\ \citenamefont
		{Peierls}(1937)}]{Mott_1937}%
	\BibitemOpen
	\bibfield  {author} {\bibinfo {author} {\bibfnamefont {N.~F.}\ \bibnamefont
			{Mott}} and \bibinfo {author} {\bibfnamefont {R.}~\bibnamefont {Peierls}},\
	}\enquote{\bibinfo {title} {Discussion of the paper by de Boer and Verwey}},\
	\href {\doibase 10.1088/0959-5309/49/4S/308} {\bibfield  {journal} {\bibinfo
			{journal} {Proceedings of the Physical Society}\ }\textbf {\bibinfo {volume}
			{49}},\ \bibinfo {pages} {72} (\bibinfo {year} {1937})}\BibitemShut {NoStop}%
	\bibitem [{\citenamefont {Mott}(1949)}]{Mott_1949}%
	\BibitemOpen
	\bibfield  {author} {\bibinfo {author} {\bibfnamefont {N.~F.}\ \bibnamefont
			{Mott}},\ }\enquote{\bibinfo {title} {The Basis of the Electron Theory of
			Metals, with Special Reference to the Transition Metals}},\ \href {\doibase
		10.1088/0370-1298/62/7/303} {\bibfield  {journal} {\bibinfo  {journal}
			{Proceedings of the Physical Society. Section A}\ }\textbf {\bibinfo {volume}
			{62}},\ \bibinfo {pages} {416} (\bibinfo {year} {1949})}\BibitemShut
	{NoStop}%
	\bibitem [{\citenamefont {Fisher}(1990)}]{Fisher:90}%
	\BibitemOpen
	\bibfield  {author} {\bibinfo {author} {\bibfnamefont {M.~P.~A.}\
			\bibnamefont {Fisher}},\ }\enquote{\bibinfo {title} {Quantum phase
			transitions in disordered two-dimensional superconductors}},\ \href {\doibase
		10.1103/PhysRevLett.65.923} {\bibfield  {journal} {\bibinfo  {journal} {Phys.
				Rev. Lett.}\ }\textbf {\bibinfo {volume} {65}},\ \bibinfo {pages} {923}
		(\bibinfo {year} {1990})}\BibitemShut {NoStop}%
	\bibitem [{\citenamefont {Sachdev}(2000)}]{Sachdev:00}%
	\BibitemOpen
	\bibfield  {author} {\bibinfo {author} {\bibfnamefont {S.}~\bibnamefont
			{Sachdev}},\ }\enquote{\bibinfo {title} {Quantum Criticality: Competing
			Ground States in Low Dimensions}},\ \href {\doibase
		10.1126/science.288.5465.475} {\bibfield  {journal} {\bibinfo  {journal}
			{Science}\ }\textbf {\bibinfo {volume} {288}},\ \bibinfo {pages} {475}
		(\bibinfo {year} {2000})}\BibitemShut {NoStop}%
	\bibitem [{\citenamefont {Fisher}(1999)}]{Fisher:99}%
	\BibitemOpen
	\bibfield  {author} {\bibinfo {author} {\bibfnamefont {D.~S.}\ \bibnamefont
			{Fisher}},\ }\enquote{\bibinfo {title} {Phase transitions and singularities
			in random quantum systems}},\ \href {\doibase
		https://doi.org/10.1016/S0378-4371(98)00498-1} {\bibfield  {journal}
		{\bibinfo  {journal} {Physica A: Statistical Mechanics and its Applications}\
		}\textbf {\bibinfo {volume} {263}},\ \bibinfo {pages} {222} (\bibinfo {year}
		{1999})},\ \bibinfo {note} {proceedings of the 20th IUPAP International
		Conference on Statistical Physics}\BibitemShut {NoStop}%
	\bibitem [{\citenamefont {Ardonne}\ \textit {et~al.}(2004)\citenamefont
		{Ardonne}, \citenamefont {Fendley},\ and\ \citenamefont {Fradkin}}]{Eddy:04}%
	\BibitemOpen
	\bibfield  {author} {\bibinfo {author} {\bibfnamefont {E.}~\bibnamefont
			{Ardonne}}, \bibinfo {author} {\bibfnamefont {P.}~\bibnamefont {Fendley}},
		and \bibinfo {author} {\bibfnamefont {E.}~\bibnamefont {Fradkin}},\
	}\enquote{\bibinfo {title} {Topological order and conformal quantum critical
			points}},\ \href {\doibase https://doi.org/10.1016/j.aop.2004.01.004}
	{\bibfield  {journal} {\bibinfo  {journal} {Annals of Physics}\ }\textbf
		{\bibinfo {volume} {310}},\ \bibinfo {pages} {493} (\bibinfo {year}
		{2004})}\BibitemShut {NoStop}%
	\bibitem [{\citenamefont {Polkovnikov}(2005)}]{Anatoli:05}%
	\BibitemOpen
	\bibfield  {author} {\bibinfo {author} {\bibfnamefont {A.}~\bibnamefont
			{Polkovnikov}},\ }\enquote{\bibinfo {title} {Universal adiabatic dynamics in
			the vicinity of a quantum critical point}},\ \href {\doibase
		10.1103/PhysRevB.72.161201} {\bibfield  {journal} {\bibinfo  {journal} {Phys.
				Rev. B}\ }\textbf {\bibinfo {volume} {72}},\ \bibinfo {pages} {161201}
		(\bibinfo {year} {2005})}\BibitemShut {NoStop}%
	\bibitem [{\citenamefont {Kist}\ \textit {et~al.}(2021)\citenamefont {Kist},
		\citenamefont {Lado},\ and\ \citenamefont {Flindt}}]{Kist:21}%
	\BibitemOpen
	\bibfield  {author} {\bibinfo {author} {\bibfnamefont {T.}~\bibnamefont
			{Kist}}, \bibinfo {author} {\bibfnamefont {J.~L.}\ \bibnamefont {Lado}},  and
		\bibinfo {author} {\bibfnamefont {C.}~\bibnamefont {Flindt}},\
	}\enquote{\bibinfo {title} {Lee-Yang theory of criticality in interacting
			quantum many-body systems}},\ \href {\doibase
		10.1103/PhysRevResearch.3.033206} {\bibfield  {journal} {\bibinfo  {journal}
			{Phys. Rev. Res.}\ }\textbf {\bibinfo {volume} {3}},\ \bibinfo {pages}
		{033206} (\bibinfo {year} {2021})}\BibitemShut {NoStop}%
	\bibitem [{\citenamefont {Osterloh}\ \textit {et~al.}(2002)\citenamefont
		{Osterloh}, \citenamefont {Amico}, \citenamefont {Falci},\ and\ \citenamefont
		{Fazio}}]{Osterloh:02}%
	\BibitemOpen
	\bibfield  {author} {\bibinfo {author} {\bibfnamefont {A.}~\bibnamefont
			{Osterloh}}, \bibinfo {author} {\bibfnamefont {L.}~\bibnamefont {Amico}},
		\bibinfo {author} {\bibfnamefont {G.}~\bibnamefont {Falci}},  and \bibinfo
		{author} {\bibfnamefont {R.}~\bibnamefont {Fazio}},\ }\enquote{\bibinfo
		{title} {Scaling of entanglement close to a quantum phase transition}},\
	\href {\doibase https://doi.org/10.1038/416608a} {\bibfield  {journal}
		{\bibinfo  {journal} {Nature}\ }\textbf {\bibinfo {volume} {416}},\ \bibinfo
		{pages} {608} (\bibinfo {year} {2002})}\BibitemShut {NoStop}%
	\bibitem [{\citenamefont {Canella}\ and\ \citenamefont
		{Fran\ifmmode~\mbox{\c{c}}\else \c{c}\fi{}a}(2021)}]{Canella:21}%
	\BibitemOpen
	\bibfield  {author} {\bibinfo {author} {\bibfnamefont {G.~A.}\ \bibnamefont
			{Canella}} and \bibinfo {author} {\bibfnamefont {V.~V.}\ \bibnamefont
			{Fran\ifmmode~\mbox{\c{c}}\else \c{c}\fi{}a}},\ }\enquote{\bibinfo {title}
		{Mott-Anderson metal-insulator transitions from entanglement}},\ \href
	{\doibase 10.1103/PhysRevB.104.134201} {\bibfield  {journal} {\bibinfo
			{journal} {Phys. Rev. B}\ }\textbf {\bibinfo {volume} {104}},\ \bibinfo
		{pages} {134201} (\bibinfo {year} {2021})}\BibitemShut {NoStop}%
	\bibitem [{\citenamefont {Canella}\ and\ \citenamefont
		{Fran{\c{c}}a}(2019)}]{Canella:19}%
	\BibitemOpen
	\bibfield  {author} {\bibinfo {author} {\bibfnamefont {G.~A.}\ \bibnamefont
			{Canella}} and \bibinfo {author} {\bibfnamefont {V.~V.}\ \bibnamefont
			{Fran{\c{c}}a}},\ }\enquote{\bibinfo {title} {Superfluid-Insulator Transition
			unambiguously detected by entanglement in one-dimensional disordered
			superfluids}},\ \href {\doibase https://doi.org/10.1038/s41598-019-51986-0}
	{\bibfield  {journal} {\bibinfo  {journal} {Scientific reports}\ }\textbf
		{\bibinfo {volume} {9}},\ \bibinfo {pages} {15313} (\bibinfo {year}
		{2019})}\BibitemShut {NoStop}%
	\bibitem [{\citenamefont {Canella}\ and\ \citenamefont
		{França}(2020)}]{Canella:20}%
	\BibitemOpen
	\bibfield  {author} {\bibinfo {author} {\bibfnamefont {G.}~\bibnamefont
			{Canella}} and \bibinfo {author} {\bibfnamefont {V.}~\bibnamefont
			{França}},\ }\enquote{\bibinfo {title} {Entanglement in disordered
			superfluids: The impact of density, interaction and harmonic confinement on
			the Superconductor–Insulator transition}},\ \href {\doibase
		https://doi.org/10.1016/j.physa.2019.123646} {\bibfield  {journal} {\bibinfo
			{journal} {Physica A: Statistical Mechanics and its Applications}\ }\textbf
		{\bibinfo {volume} {545}},\ \bibinfo {pages} {123646} (\bibinfo {year}
		{2020})}\BibitemShut {NoStop}%
	\bibitem [{\citenamefont {Osborne}\ and\ \citenamefont
		{Nielsen}(2002)}]{Osborne:02}%
	\BibitemOpen
	\bibfield  {author} {\bibinfo {author} {\bibfnamefont {T.~J.}\ \bibnamefont
			{Osborne}} and \bibinfo {author} {\bibfnamefont {M.~A.}\ \bibnamefont
			{Nielsen}},\ }\enquote{\bibinfo {title} {Entanglement in a simple quantum
			phase transition}},\ \href {\doibase 10.1103/PhysRevA.66.032110} {\bibfield
		{journal} {\bibinfo  {journal} {Phys. Rev. A}\ }\textbf {\bibinfo {volume}
			{66}},\ \bibinfo {pages} {032110} (\bibinfo {year} {2002})}\BibitemShut
	{NoStop}%
	\bibitem [{\citenamefont {Vidal}\ \textit {et~al.}(2003)\citenamefont {Vidal},
		\citenamefont {Latorre}, \citenamefont {Rico},\ and\ \citenamefont
		{Kitaev}}]{Vidal:03}%
	\BibitemOpen
	\bibfield  {author} {\bibinfo {author} {\bibfnamefont {G.}~\bibnamefont
			{Vidal}}, \bibinfo {author} {\bibfnamefont {J.~I.}\ \bibnamefont {Latorre}},
		\bibinfo {author} {\bibfnamefont {E.}~\bibnamefont {Rico}},  and \bibinfo
		{author} {\bibfnamefont {A.}~\bibnamefont {Kitaev}},\ }\enquote{\bibinfo
		{title} {Entanglement in Quantum Critical Phenomena}},\ \href {\doibase
		10.1103/PhysRevLett.90.227902} {\bibfield  {journal} {\bibinfo  {journal}
			{Phys. Rev. Lett.}\ }\textbf {\bibinfo {volume} {90}},\ \bibinfo {pages}
		{227902} (\bibinfo {year} {2003})}\BibitemShut {NoStop}%
	\bibitem [{\citenamefont {Verstraete}\ \textit {et~al.}(2004)\citenamefont
		{Verstraete}, \citenamefont {Popp},\ and\ \citenamefont
		{Cirac}}]{Verstraete:04}%
	\BibitemOpen
	\bibfield  {author} {\bibinfo {author} {\bibfnamefont {F.}~\bibnamefont
			{Verstraete}}, \bibinfo {author} {\bibfnamefont {M.}~\bibnamefont {Popp}},
		and \bibinfo {author} {\bibfnamefont {J.~I.}\ \bibnamefont {Cirac}},\
	}\enquote{\bibinfo {title} {Entanglement versus Correlations in Spin
			Systems}},\ \href {\doibase 10.1103/PhysRevLett.92.027901} {\bibfield
		{journal} {\bibinfo  {journal} {Phys. Rev. Lett.}\ }\textbf {\bibinfo
			{volume} {92}},\ \bibinfo {pages} {027901} (\bibinfo {year}
		{2004})}\BibitemShut {NoStop}%
	\bibitem [{\citenamefont {Wu}\ \textit {et~al.}(2004)\citenamefont {Wu},
		\citenamefont {Sarandy},\ and\ \citenamefont {Lidar}}]{Wu:04}%
	\BibitemOpen
	\bibfield  {author} {\bibinfo {author} {\bibfnamefont {L.-A.}\ \bibnamefont
			{Wu}}, \bibinfo {author} {\bibfnamefont {M.~S.}\ \bibnamefont {Sarandy}},
		and \bibinfo {author} {\bibfnamefont {D.~A.}\ \bibnamefont {Lidar}},\
	}\enquote{\bibinfo {title} {Quantum Phase Transitions and Bipartite
			Entanglement}},\ \href {\doibase 10.1103/PhysRevLett.93.250404} {\bibfield
		{journal} {\bibinfo  {journal} {Phys. Rev. Lett.}\ }\textbf {\bibinfo
			{volume} {93}},\ \bibinfo {pages} {250404} (\bibinfo {year}
		{2004})}\BibitemShut {NoStop}%
	\bibitem [{\citenamefont {Gu}\ \textit {et~al.}(2004)\citenamefont {Gu},
		\citenamefont {Deng}, \citenamefont {Li},\ and\ \citenamefont {Lin}}]{Gu:04}%
	\BibitemOpen
	\bibfield  {author} {\bibinfo {author} {\bibfnamefont {S.-J.}\ \bibnamefont
			{Gu}}, \bibinfo {author} {\bibfnamefont {S.-S.}\ \bibnamefont {Deng}},
		\bibinfo {author} {\bibfnamefont {Y.-Q.}\ \bibnamefont {Li}},  and \bibinfo
		{author} {\bibfnamefont {H.-Q.}\ \bibnamefont {Lin}},\ }\enquote{\bibinfo
		{title} {Entanglement and Quantum Phase Transition in the Extended Hubbard
			Model}},\ \href {\doibase 10.1103/PhysRevLett.93.086402} {\bibfield
		{journal} {\bibinfo  {journal} {Phys. Rev. Lett.}\ }\textbf {\bibinfo
			{volume} {93}},\ \bibinfo {pages} {086402} (\bibinfo {year}
		{2004})}\BibitemShut {NoStop}%
	\bibitem [{\citenamefont {Larsson}\ and\ \citenamefont
		{Johannesson}(2005)}]{Larsson:05}%
	\BibitemOpen
	\bibfield  {author} {\bibinfo {author} {\bibfnamefont {D.}~\bibnamefont
			{Larsson}} and \bibinfo {author} {\bibfnamefont {H.}~\bibnamefont
			{Johannesson}},\ }\enquote{\bibinfo {title} {Entanglement Scaling in the
			One-Dimensional Hubbard Model at Criticality}},\ \href {\doibase
		10.1103/PhysRevLett.95.196406} {\bibfield  {journal} {\bibinfo  {journal}
			{Phys. Rev. Lett.}\ }\textbf {\bibinfo {volume} {95}},\ \bibinfo {pages}
		{196406} (\bibinfo {year} {2005})}\BibitemShut {NoStop}%
	\bibitem [{\citenamefont {Sarandy}(2009)}]{Sarandy:09}%
	\BibitemOpen
	\bibfield  {author} {\bibinfo {author} {\bibfnamefont {M.~S.}\ \bibnamefont
			{Sarandy}},\ }\enquote{\bibinfo {title} {Classical correlation and quantum
			discord in critical systems}},\ \href {\doibase 10.1103/PhysRevA.80.022108}
	{\bibfield  {journal} {\bibinfo  {journal} {Phys. Rev. A}\ }\textbf {\bibinfo
			{volume} {80}},\ \bibinfo {pages} {022108} (\bibinfo {year}
		{2009})}\BibitemShut {NoStop}%
	\bibitem [{\citenamefont {Dillenschneider}(2008)}]{Dillenschneider:08}%
	\BibitemOpen
	\bibfield  {author} {\bibinfo {author} {\bibfnamefont {R.}~\bibnamefont
			{Dillenschneider}},\ }\enquote{\bibinfo {title} {Quantum discord and quantum
			phase transition in spin chains}},\ \href {\doibase
		10.1103/PhysRevB.78.224413} {\bibfield  {journal} {\bibinfo  {journal} {Phys.
				Rev. B}\ }\textbf {\bibinfo {volume} {78}},\ \bibinfo {pages} {224413}
		(\bibinfo {year} {2008})}\BibitemShut {NoStop}%
	\bibitem [{\citenamefont {Hur}(2008)}]{HUR20082208}%
	\BibitemOpen
	\bibfield  {author} {\bibinfo {author} {\bibfnamefont {K.~L.}\ \bibnamefont
			{Hur}},\ }\enquote{\bibinfo {title} {Entanglement entropy, decoherence, and
			quantum phase transitions of a dissipative two-level system}},\ \href
	{\doibase https://doi.org/10.1016/j.aop.2007.12.003} {\bibfield  {journal}
		{\bibinfo  {journal} {Annals of Physics}\ }\textbf {\bibinfo {volume}
			{323}},\ \bibinfo {pages} {2208} (\bibinfo {year} {2008})}\BibitemShut
	{NoStop}%
	\bibitem [{\citenamefont {Jia}\ \textit {et~al.}(2008)\citenamefont {Jia},
		\citenamefont {Subramaniam}, \citenamefont {Gruzberg},\ and\ \citenamefont
		{Chakravarty}}]{Jia:08}%
	\BibitemOpen
	\bibfield  {author} {\bibinfo {author} {\bibfnamefont {X.}~\bibnamefont
			{Jia}}, \bibinfo {author} {\bibfnamefont {A.~R.}\ \bibnamefont
			{Subramaniam}}, \bibinfo {author} {\bibfnamefont {I.~A.}\ \bibnamefont
			{Gruzberg}},  and \bibinfo {author} {\bibfnamefont {S.}~\bibnamefont
			{Chakravarty}},\ }\enquote{\bibinfo {title} {Entanglement entropy and
			multifractality at localization transitions}},\ \href {\doibase
		10.1103/PhysRevB.77.014208} {\bibfield  {journal} {\bibinfo  {journal} {Phys.
				Rev. B}\ }\textbf {\bibinfo {volume} {77}},\ \bibinfo {pages} {014208}
		(\bibinfo {year} {2008})}\BibitemShut {NoStop}%
	\bibitem [{\citenamefont {Coulamy}\ \textit {et~al.}(2017)\citenamefont
		{Coulamy}, \citenamefont {Saguia},\ and\ \citenamefont {Sarandy}}]{Ivan:17}%
	\BibitemOpen
	\bibfield  {author} {\bibinfo {author} {\bibfnamefont {I.~B.}\ \bibnamefont
			{Coulamy}}, \bibinfo {author} {\bibfnamefont {A.}~\bibnamefont {Saguia}},
		and \bibinfo {author} {\bibfnamefont {M.~S.}\ \bibnamefont {Sarandy}},\
	}\enquote{\bibinfo {title} {Dynamics of the quantum search and quench-induced
			first-order phase transitions}},\ \href {\doibase 10.1103/PhysRevE.95.022127}
	{\bibfield  {journal} {\bibinfo  {journal} {Phys. Rev. E}\ }\textbf {\bibinfo
			{volume} {95}},\ \bibinfo {pages} {022127} (\bibinfo {year}
		{2017})}\BibitemShut {NoStop}%
	\bibitem [{\citenamefont {Filho}\ \textit {et~al.}(2022)\citenamefont {Filho},
		\citenamefont {Izquierdo}, \citenamefont {Saguia}, \citenamefont {Albash},
		\citenamefont {Hen},\ and\ \citenamefont {Sarandy}}]{Jaime:22}%
	\BibitemOpen
	\bibfield  {author} {\bibinfo {author} {\bibfnamefont {J.~L. C. d.~C.}\
			\bibnamefont {Filho}}, \bibinfo {author} {\bibfnamefont {Z.~G.}\ \bibnamefont
			{Izquierdo}}, \bibinfo {author} {\bibfnamefont {A.}~\bibnamefont {Saguia}},
		\bibinfo {author} {\bibfnamefont {T.}~\bibnamefont {Albash}}, \bibinfo
		{author} {\bibfnamefont {I.}~\bibnamefont {Hen}},  and \bibinfo {author}
		{\bibfnamefont {M.~S.}\ \bibnamefont {Sarandy}},\ }\enquote{\bibinfo {title}
		{Localization transition induced by programmable disorder}},\ \href {\doibase
		10.1103/PhysRevB.105.134201} {\bibfield  {journal} {\bibinfo  {journal}
			{Phys. Rev. B}\ }\textbf {\bibinfo {volume} {105}},\ \bibinfo {pages}
		{134201} (\bibinfo {year} {2022})}\BibitemShut {NoStop}%
	\bibitem [{\citenamefont {Kopp}\ \textit {et~al.}(2007)\citenamefont {Kopp},
		\citenamefont {Jia},\ and\ \citenamefont {Chakravarty}}]{KOPP20071466}%
	\BibitemOpen
	\bibfield  {author} {\bibinfo {author} {\bibfnamefont {A.}~\bibnamefont
			{Kopp}}, \bibinfo {author} {\bibfnamefont {X.}~\bibnamefont {Jia}},  and
		\bibinfo {author} {\bibfnamefont {S.}~\bibnamefont {Chakravarty}},\
	}\enquote{\bibinfo {title} {Replacing energy by von Neumann entropy in
			quantum phase transitions}},\ \href {\doibase
		https://doi.org/10.1016/j.aop.2006.08.002} {\bibfield  {journal} {\bibinfo
			{journal} {Annals of Physics}\ }\textbf {\bibinfo {volume} {322}},\ \bibinfo
		{pages} {1466} (\bibinfo {year} {2007})}\BibitemShut {NoStop}%
	\bibitem [{\citenamefont {Wicks}\ \textit {et~al.}(2007)\citenamefont {Wicks},
		\citenamefont {Chapman},\ and\ \citenamefont {Dendy}}]{Wicks:07}%
	\BibitemOpen
	\bibfield  {author} {\bibinfo {author} {\bibfnamefont {R.~T.}\ \bibnamefont
			{Wicks}}, \bibinfo {author} {\bibfnamefont {S.~C.}\ \bibnamefont {Chapman}},
		and \bibinfo {author} {\bibfnamefont {R.~O.}\ \bibnamefont {Dendy}},\
	}\enquote{\bibinfo {title} {Mutual information as a tool for identifying
			phase transitions in dynamical complex systems with limited data}},\ \href
	{\doibase 10.1103/PhysRevE.75.051125} {\bibfield  {journal} {\bibinfo
			{journal} {Phys. Rev. E}\ }\textbf {\bibinfo {volume} {75}},\ \bibinfo
		{pages} {051125} (\bibinfo {year} {2007})}\BibitemShut {NoStop}%
	\bibitem [{\citenamefont {Chen}\ and\ \citenamefont {Li}(2010)}]{Chen:10}%
	\BibitemOpen
	\bibfield  {author} {\bibinfo {author} {\bibfnamefont {Y.-X.}\ \bibnamefont
			{Chen}} and \bibinfo {author} {\bibfnamefont {S.-W.}\ \bibnamefont {Li}},\
	}\enquote{\bibinfo {title} {Quantum correlations in topological quantum phase
			transitions}},\ \href {\doibase 10.1103/PhysRevA.81.032120} {\bibfield
		{journal} {\bibinfo  {journal} {Phys. Rev. A}\ }\textbf {\bibinfo {volume}
			{81}},\ \bibinfo {pages} {032120} (\bibinfo {year} {2010})}\BibitemShut
	{NoStop}%
	\bibitem [{\citenamefont {Dong}\ \textit {et~al.}(2021)\citenamefont {Dong},
		\citenamefont {Huang},\ and\ \citenamefont {Yang}}]{Dong:21}%
	\BibitemOpen
	\bibfield  {author} {\bibinfo {author} {\bibfnamefont {J.-J.}\ \bibnamefont
			{Dong}}, \bibinfo {author} {\bibfnamefont {D.}~\bibnamefont {Huang}},  and
		\bibinfo {author} {\bibfnamefont {Y.-f.}\ \bibnamefont {Yang}},\
	}\enquote{\bibinfo {title} {Mutual information, quantum phase transition, and
			phase coherence in Kondo systems}},\ \href {\doibase
		10.1103/PhysRevB.104.L081115} {\bibfield  {journal} {\bibinfo  {journal}
			{Phys. Rev. B}\ }\textbf {\bibinfo {volume} {104}},\ \bibinfo {pages}
		{L081115} (\bibinfo {year} {2021})}\BibitemShut {NoStop}%
	\bibitem [{\citenamefont {Milne}\ \textit {et~al.}(2014)\citenamefont {Milne},
		\citenamefont {Jevtic}, \citenamefont {Jennings}, \citenamefont {Wiseman},\
		and\ \citenamefont {Rudolph}}]{Milne_2014}%
	\BibitemOpen
	\bibfield  {author} {\bibinfo {author} {\bibfnamefont {A.}~\bibnamefont
			{Milne}}, \bibinfo {author} {\bibfnamefont {S.}~\bibnamefont {Jevtic}},
		\bibinfo {author} {\bibfnamefont {D.}~\bibnamefont {Jennings}}, \bibinfo
		{author} {\bibfnamefont {H.}~\bibnamefont {Wiseman}},  and \bibinfo {author}
		{\bibfnamefont {T.}~\bibnamefont {Rudolph}},\ }\enquote{\bibinfo {title}
		{Quantum steering ellipsoids, extremal physical states and monogamy}},\ \href
	{\doibase 10.1088/1367-2630/16/8/083017} {\bibfield  {journal} {\bibinfo
			{journal} {New Journal of Physics}\ }\textbf {\bibinfo {volume} {16}},\
		\bibinfo {pages} {083017} (\bibinfo {year} {2014})}\BibitemShut {NoStop}%
	\bibitem [{\citenamefont {Jevtic}\ \textit {et~al.}(2014)\citenamefont
		{Jevtic}, \citenamefont {Pusey}, \citenamefont {Jennings},\ and\
		\citenamefont {Rudolph}}]{Jevtic_2014}%
	\BibitemOpen
	\bibfield  {author} {\bibinfo {author} {\bibfnamefont {S.}~\bibnamefont
			{Jevtic}}, \bibinfo {author} {\bibfnamefont {M.}~\bibnamefont {Pusey}},
		\bibinfo {author} {\bibfnamefont {D.}~\bibnamefont {Jennings}},  and \bibinfo
		{author} {\bibfnamefont {T.}~\bibnamefont {Rudolph}},\ }\enquote{\bibinfo
		{title} {Quantum Steering Ellipsoids}},\ \href {\doibase
		10.1103/PhysRevLett.113.020402} {\bibfield  {journal} {\bibinfo  {journal}
			{Phys. Rev. Lett.}\ }\textbf {\bibinfo {volume} {113}},\ \bibinfo {pages}
		{020402} (\bibinfo {year} {2014})}\BibitemShut {NoStop}%
	\bibitem [{\citenamefont {Xu}\ \textit {et~al.}(2023)\citenamefont {Xu},
		\citenamefont {Liu}, \citenamefont {Wang}, \citenamefont {Zhang},
		\citenamefont {Huang}, \citenamefont {Liu}, \citenamefont {Cheng},
		\citenamefont {Li},\ and\ \citenamefont {Guo}}]{xu2023experimental}%
	\BibitemOpen
	\bibfield  {author} {\bibinfo {author} {\bibfnamefont {K.}~\bibnamefont
			{Xu}}, \bibinfo {author} {\bibfnamefont {L.}~\bibnamefont {Liu}}, \bibinfo
		{author} {\bibfnamefont {N.-N.}\ \bibnamefont {Wang}}, \bibinfo {author}
		{\bibfnamefont {C.}~\bibnamefont {Zhang}}, \bibinfo {author} {\bibfnamefont
			{Y.-F.}\ \bibnamefont {Huang}}, \bibinfo {author} {\bibfnamefont {B.-H.}\
			\bibnamefont {Liu}}, \bibinfo {author} {\bibfnamefont {S.}~\bibnamefont
			{Cheng}}, \bibinfo {author} {\bibfnamefont {C.-F.}\ \bibnamefont {Li}},  and
		\bibinfo {author} {\bibfnamefont {G.-C.}\ \bibnamefont {Guo}},\ }\href@noop
	{} {\enquote{\bibinfo {title} {Experimental verification of the steering
				ellipsoid zoo via two-qubit states}}} (\bibinfo {year} {2023}),\ \Eprint
	{http://arxiv.org/abs/2310.18645} {arXiv:2310.18645 [quant-ph]} \BibitemShut
	{NoStop}%
	\bibitem [{\citenamefont {Rosario}\ \textit {et~al.}(2023)\citenamefont
		{Rosario}, \citenamefont {Ducuara},\ and\ \citenamefont
		{Susa}}]{Rosario_2023}%
	\BibitemOpen
	\bibfield  {author} {\bibinfo {author} {\bibfnamefont {P.}~\bibnamefont
			{Rosario}}, \bibinfo {author} {\bibfnamefont {A.~F.}\ \bibnamefont
			{Ducuara}},  and \bibinfo {author} {\bibfnamefont {C.~E.}\ \bibnamefont
			{Susa}},\ }\href@noop {} {\enquote{\bibinfo {title} {Swapping of quantum
				correlations and the role of local filtering operations}}} (\bibinfo {year}
	{2023}),\ \Eprint {http://arxiv.org/abs/2307.16524} {arXiv:2307.16524
		[quant-ph]} \BibitemShut {NoStop}%
	\bibitem [{\citenamefont {Du}\ \textit {et~al.}(2021)\citenamefont {Du},
		\citenamefont {Zhang}, \citenamefont {Zhou},\ and\ \citenamefont
		{Tong}}]{Du:21}%
	\BibitemOpen
	\bibfield  {author} {\bibinfo {author} {\bibfnamefont {M.-M.}\ \bibnamefont
			{Du}}, \bibinfo {author} {\bibfnamefont {D.-J.}\ \bibnamefont {Zhang}},
		\bibinfo {author} {\bibfnamefont {Z.-Y.}\ \bibnamefont {Zhou}},  and \bibinfo
		{author} {\bibfnamefont {D.~M.}\ \bibnamefont {Tong}},\ }\enquote{\bibinfo
		{title} {Visualizing quantum phase transitions in the $XXZ$ model via the
			quantum steering ellipsoid}},\ \href {\doibase 10.1103/PhysRevA.104.012418}
	{\bibfield  {journal} {\bibinfo  {journal} {Phys. Rev. A}\ }\textbf {\bibinfo
			{volume} {104}},\ \bibinfo {pages} {012418} (\bibinfo {year}
		{2021})}\BibitemShut {NoStop}%
	\bibitem [{\citenamefont {Gisin}(1996)}]{Gisin:96}%
	\BibitemOpen
	\bibfield  {author} {\bibinfo {author} {\bibfnamefont {N.}~\bibnamefont
			{Gisin}},\ }\enquote{\bibinfo {title} {Hidden quantum nonlocality revealed by
			local filters}},\ \href {\doibase
		https://doi.org/10.1016/S0375-9601(96)80001-6} {\bibfield  {journal}
		{\bibinfo  {journal} {Physics Letters A}\ }\textbf {\bibinfo {volume}
			{210}},\ \bibinfo {pages} {151} (\bibinfo {year} {1996})}\BibitemShut
	{NoStop}%
	\bibitem [{\citenamefont {Verstraete}\ \textit {et~al.}(2001)\citenamefont
		{Verstraete}, \citenamefont {Dehaene},\ and\ \citenamefont
		{DeMoor}}]{Verstraete:01}%
	\BibitemOpen
	\bibfield  {author} {\bibinfo {author} {\bibfnamefont {F.}~\bibnamefont
			{Verstraete}}, \bibinfo {author} {\bibfnamefont {J.}~\bibnamefont {Dehaene}},
		and \bibinfo {author} {\bibfnamefont {B.}~\bibnamefont {DeMoor}},\
	}\enquote{\bibinfo {title} {Local filtering operations on two qubits}},\
	\href {\doibase 10.1103/PhysRevA.64.010101} {\bibfield  {journal} {\bibinfo
			{journal} {Phys. Rev. A}\ }\textbf {\bibinfo {volume} {64}},\ \bibinfo
		{pages} {010101} (\bibinfo {year} {2001})}\BibitemShut {NoStop}%
	\bibitem [{\citenamefont {Verstraete}\ and\ \citenamefont
		{Wolf}(2002)}]{Verstraete:02}%
	\BibitemOpen
	\bibfield  {author} {\bibinfo {author} {\bibfnamefont {F.}~\bibnamefont
			{Verstraete}} and \bibinfo {author} {\bibfnamefont {M.~M.}\ \bibnamefont
			{Wolf}},\ }\enquote{\bibinfo {title} {Entanglement versus Bell Violations and
			Their Behavior under Local Filtering Operations}},\ \href {\doibase
		10.1103/PhysRevLett.89.170401} {\bibfield  {journal} {\bibinfo  {journal}
			{Phys. Rev. Lett.}\ }\textbf {\bibinfo {volume} {89}},\ \bibinfo {pages}
		{170401} (\bibinfo {year} {2002})}\BibitemShut {NoStop}%
	\bibitem [{\citenamefont {Nielsen}\ and\ \citenamefont
		{Chuang}(2011)}]{Nielsen:Book}%
	\BibitemOpen
	\bibfield  {author} {\bibinfo {author} {\bibfnamefont {M.~A.}\ \bibnamefont
			{Nielsen}} and \bibinfo {author} {\bibfnamefont {I.~L.}\ \bibnamefont
			{Chuang}},\ }\href {\doibase 10.1017/CBO9780511976667} {\textit {\bibinfo
			{title} {Quantum Computation and Quantum Information: 10th Anniversary
				Edition}}},\ \bibinfo {edition} {10th}\ ed.\ (\bibinfo  {publisher}
	{Cambridge University Press},\ \bibinfo {address} {New York, NY, USA},\
	\bibinfo {year} {2011})\BibitemShut {NoStop}%
	\bibitem [{\citenamefont {Gamel}(2016)}]{Gamel_2016}%
	\BibitemOpen
	\bibfield  {author} {\bibinfo {author} {\bibfnamefont {O.}~\bibnamefont
			{Gamel}},\ }\enquote{\bibinfo {title} {Entangled Bloch spheres: Bloch matrix
			and two-qubit state space}},\ \href {\doibase 10.1103/PhysRevA.93.062320}
	{\bibfield  {journal} {\bibinfo  {journal} {Phys. Rev. A}\ }\textbf {\bibinfo
			{volume} {93}},\ \bibinfo {pages} {062320} (\bibinfo {year}
		{2016})}\BibitemShut {NoStop}%
	\bibitem [{\citenamefont {Horodecki}\ \textit {et~al.}(1995)\citenamefont
		{Horodecki}, \citenamefont {Horodecki},\ and\ \citenamefont
		{Horodecki}}]{Horodecki_1995_CHSH}%
	\BibitemOpen
	\bibfield  {author} {\bibinfo {author} {\bibfnamefont {R.}~\bibnamefont
			{Horodecki}}, \bibinfo {author} {\bibfnamefont {P.}~\bibnamefont
			{Horodecki}},  and \bibinfo {author} {\bibfnamefont {M.}~\bibnamefont
			{Horodecki}},\ }\enquote{\bibinfo {title} {Violating Bell inequality by mixed
			spin-12 states: necessary and sufficient condition}},\ \href {\doibase
		https://doi.org/10.1016/0375-9601(95)00214-N} {\bibfield  {journal} {\bibinfo
			{journal} {Physics Letters A}\ }\textbf {\bibinfo {volume} {200}},\ \bibinfo
		{pages} {340} (\bibinfo {year} {1995})}\BibitemShut {NoStop}%
	\bibitem [{\citenamefont {Costa}\ and\ \citenamefont
		{Angelo}(2016)}]{Costa_2016}%
	\BibitemOpen
	\bibfield  {author} {\bibinfo {author} {\bibfnamefont {A.~C.~S.}\
			\bibnamefont {Costa}} and \bibinfo {author} {\bibfnamefont {R.~M.}\
			\bibnamefont {Angelo}},\ }\enquote{\bibinfo {title} {Quantification of
			Einstein-Podolsky-Rosen steering for two-qubit states}},\ \href {\doibase
		10.1103/PhysRevA.93.020103} {\bibfield  {journal} {\bibinfo  {journal} {Phys.
				Rev. A}\ }\textbf {\bibinfo {volume} {93}},\ \bibinfo {pages} {020103}
		(\bibinfo {year} {2016})}\BibitemShut {NoStop}%
	\bibitem [{\citenamefont {Horodecki}\ \textit {et~al.}(1996)\citenamefont
		{Horodecki}, \citenamefont {Horodecki},\ and\ \citenamefont
		{Horodecki}}]{Horodecki:96}%
	\BibitemOpen
	\bibfield  {author} {\bibinfo {author} {\bibfnamefont {R.}~\bibnamefont
			{Horodecki}}, \bibinfo {author} {\bibfnamefont {M.}~\bibnamefont
			{Horodecki}},  and \bibinfo {author} {\bibfnamefont {P.}~\bibnamefont
			{Horodecki}},\ }\enquote{\bibinfo {title} {Teleportation, Bell's inequalities
			and inseparability}},\ \href {\doibase
		https://doi.org/10.1016/0375-9601(96)00639-1} {\bibfield  {journal} {\bibinfo
			{journal} {Physics Letters A}\ }\textbf {\bibinfo {volume} {222}},\ \bibinfo
		{pages} {21} (\bibinfo {year} {1996})}\BibitemShut {NoStop}%
	\bibitem [{\citenamefont {{de Gennes}}(1963)}]{deGennes:63}%
	\BibitemOpen
	\bibfield  {author} {\bibinfo {author} {\bibfnamefont {P.}~\bibnamefont {{de
					Gennes}}},\ }\enquote{\bibinfo {title} {Collective motions of hydrogen
			bonds}},\ \href {\doibase https://doi.org/10.1016/0038-1098(63)90212-6}
	{\bibfield  {journal} {\bibinfo  {journal} {Solid State Communications}\
		}\textbf {\bibinfo {volume} {1}},\ \bibinfo {pages} {132} (\bibinfo {year}
		{1963})}\BibitemShut {NoStop}%
	\bibitem [{\citenamefont {Barouch}\ \textit {et~al.}(1970)\citenamefont
		{Barouch}, \citenamefont {McCoy},\ and\ \citenamefont
		{Dresden}}]{Barouch:70}%
	\BibitemOpen
	\bibfield  {author} {\bibinfo {author} {\bibfnamefont {E.}~\bibnamefont
			{Barouch}}, \bibinfo {author} {\bibfnamefont {B.~M.}\ \bibnamefont {McCoy}},
		and \bibinfo {author} {\bibfnamefont {M.}~\bibnamefont {Dresden}},\
	}\enquote{\bibinfo {title} {Statistical Mechanics of the $\mathrm{XY}$ Model.
			I}},\ \href {\doibase 10.1103/PhysRevA.2.1075} {\bibfield  {journal}
		{\bibinfo  {journal} {Phys. Rev. A}\ }\textbf {\bibinfo {volume} {2}},\
		\bibinfo {pages} {1075} (\bibinfo {year} {1970})}\BibitemShut {NoStop}%
	\bibitem [{\citenamefont {Barouch}\ and\ \citenamefont
		{McCoy}(1971)}]{Barouch:71}%
	\BibitemOpen
	\bibfield  {author} {\bibinfo {author} {\bibfnamefont {E.}~\bibnamefont
			{Barouch}} and \bibinfo {author} {\bibfnamefont {B.~M.}\ \bibnamefont
			{McCoy}},\ }\enquote{\bibinfo {title} {Statistical Mechanics of the $XY$
			Model. II. Spin-Correlation Functions}},\ \href {\doibase
		10.1103/PhysRevA.3.786} {\bibfield  {journal} {\bibinfo  {journal} {Phys.
				Rev. A}\ }\textbf {\bibinfo {volume} {3}},\ \bibinfo {pages} {786} (\bibinfo
		{year} {1971})}\BibitemShut {NoStop}%
	\bibitem [{\citenamefont {Maziero}\ \textit {et~al.}(2010)\citenamefont
		{Maziero}, \citenamefont {Guzman}, \citenamefont {C\'eleri}, \citenamefont
		{Sarandy},\ and\ \citenamefont {Serra}}]{Maziero:10}%
	\BibitemOpen
	\bibfield  {author} {\bibinfo {author} {\bibfnamefont {J.}~\bibnamefont
			{Maziero}}, \bibinfo {author} {\bibfnamefont {H.~C.}\ \bibnamefont {Guzman}},
		\bibinfo {author} {\bibfnamefont {L.~C.}\ \bibnamefont {C\'eleri}}, \bibinfo
		{author} {\bibfnamefont {M.~S.}\ \bibnamefont {Sarandy}},  and \bibinfo
		{author} {\bibfnamefont {R.~M.}\ \bibnamefont {Serra}},\ }\enquote{\bibinfo
		{title} {Quantum and classical thermal correlations in the $\mathit{XY}$
			spin-$\frac{1}{2}$ chain}},\ \href {\doibase 10.1103/PhysRevA.82.012106}
	{\bibfield  {journal} {\bibinfo  {journal} {Phys. Rev. A}\ }\textbf {\bibinfo
			{volume} {82}},\ \bibinfo {pages} {012106} (\bibinfo {year}
		{2010})}\BibitemShut {NoStop}%
	\bibitem [{\citenamefont {Yang}\ and\ \citenamefont
		{Yang}(1966{\natexlab{a}})}]{Yang:660}%
	\BibitemOpen
	\bibfield  {author} {\bibinfo {author} {\bibfnamefont {C.~N.}\ \bibnamefont
			{Yang}} and \bibinfo {author} {\bibfnamefont {C.~P.}\ \bibnamefont {Yang}},\
	}\enquote{\bibinfo {title} {Ground-State Energy of a Heisenberg-Ising
			Lattice}},\ \href {\doibase 10.1103/PhysRev.147.303} {\bibfield  {journal}
		{\bibinfo  {journal} {Phys. Rev.}\ }\textbf {\bibinfo {volume} {147}},\
		\bibinfo {pages} {303} (\bibinfo {year} {1966}{\natexlab{a}})}\BibitemShut
	{NoStop}%
	\bibitem [{\citenamefont {Yang}\ and\ \citenamefont
		{Yang}(1966{\natexlab{b}})}]{Yang:66}%
	\BibitemOpen
	\bibfield  {author} {\bibinfo {author} {\bibfnamefont {C.~N.}\ \bibnamefont
			{Yang}} and \bibinfo {author} {\bibfnamefont {C.~P.}\ \bibnamefont {Yang}},\
	}\enquote{\bibinfo {title} {One-Dimensional Chain of Anisotropic Spin-Spin
			Interactions. I. Proof of Bethe's Hypothesis for Ground State in a Finite
			System}},\ \href {\doibase 10.1103/PhysRev.150.321} {\bibfield  {journal}
		{\bibinfo  {journal} {Phys. Rev.}\ }\textbf {\bibinfo {volume} {150}},\
		\bibinfo {pages} {321} (\bibinfo {year} {1966}{\natexlab{b}})}\BibitemShut
	{NoStop}%
	\bibitem [{\citenamefont {Yang}\ and\ \citenamefont
		{Yang}(1966{\natexlab{c}})}]{Yang:66II}%
	\BibitemOpen
	\bibfield  {author} {\bibinfo {author} {\bibfnamefont {C.~N.}\ \bibnamefont
			{Yang}} and \bibinfo {author} {\bibfnamefont {C.~P.}\ \bibnamefont {Yang}},\
	}\enquote{\bibinfo {title} {One-Dimensional Chain of Anisotropic Spin-Spin
			Interactions. II. Properties of the Ground-State Energy Per Lattice Site for
			an Infinite System}},\ \href {\doibase 10.1103/PhysRev.150.327} {\bibfield
		{journal} {\bibinfo  {journal} {Phys. Rev.}\ }\textbf {\bibinfo {volume}
			{150}},\ \bibinfo {pages} {327} (\bibinfo {year}
		{1966}{\natexlab{c}})}\BibitemShut {NoStop}%
	\bibitem [{\citenamefont {Gersch}\ and\ \citenamefont
		{Knollman}(1963)}]{Gersch:63}%
	\BibitemOpen
	\bibfield  {author} {\bibinfo {author} {\bibfnamefont {H.~A.}\ \bibnamefont
			{Gersch}} and \bibinfo {author} {\bibfnamefont {G.~C.}\ \bibnamefont
			{Knollman}},\ }\enquote{\bibinfo {title} {Quantum Cell Model for Bosons}},\
	\href {\doibase 10.1103/PhysRev.129.959} {\bibfield  {journal} {\bibinfo
			{journal} {Phys. Rev.}\ }\textbf {\bibinfo {volume} {129}},\ \bibinfo {pages}
		{959} (\bibinfo {year} {1963})}\BibitemShut {NoStop}%
	\bibitem [{\citenamefont {Tsuchiya}\ \textit {et~al.}(2004)\citenamefont
		{Tsuchiya}, \citenamefont {Kurihara},\ and\ \citenamefont
		{Kimura}}]{Tsuchiya:04}%
	\BibitemOpen
	\bibfield  {author} {\bibinfo {author} {\bibfnamefont {S.}~\bibnamefont
			{Tsuchiya}}, \bibinfo {author} {\bibfnamefont {S.}~\bibnamefont {Kurihara}},
		and \bibinfo {author} {\bibfnamefont {T.}~\bibnamefont {Kimura}},\
	}\enquote{\bibinfo {title} {Superfluid--Mott insulator transition of spin-1
			bosons in an optical lattice}},\ \href {\doibase 10.1103/PhysRevA.70.043628}
	{\bibfield  {journal} {\bibinfo  {journal} {Phys. Rev. A}\ }\textbf {\bibinfo
			{volume} {70}},\ \bibinfo {pages} {043628} (\bibinfo {year}
		{2004})}\BibitemShut {NoStop}%
	\bibitem [{\citenamefont {Albus}\ \textit {et~al.}(2003)\citenamefont {Albus},
		\citenamefont {Illuminati},\ and\ \citenamefont {Eisert}}]{Albus:03}%
	\BibitemOpen
	\bibfield  {author} {\bibinfo {author} {\bibfnamefont {A.}~\bibnamefont
			{Albus}}, \bibinfo {author} {\bibfnamefont {F.}~\bibnamefont {Illuminati}},
		and \bibinfo {author} {\bibfnamefont {J.}~\bibnamefont {Eisert}},\
	}\enquote{\bibinfo {title} {Mixtures of bosonic and fermionic atoms in
			optical lattices}},\ \href {\doibase 10.1103/PhysRevA.68.023606} {\bibfield
		{journal} {\bibinfo  {journal} {Phys. Rev. A}\ }\textbf {\bibinfo {volume}
			{68}},\ \bibinfo {pages} {023606} (\bibinfo {year} {2003})}\BibitemShut
	{NoStop}%
\end{thebibliography}

%

\end{document}